\documentclass[reprint, aps, pre, floatfix]{revtex4-2}
\usepackage[utf8]{inputenc}
\usepackage{amsmath}
\usepackage{bm}
\usepackage{graphicx}
\usepackage{mathptmx}
\usepackage{amsthm}
\newtheorem{theorem}{Theorem}
\newtheorem{lemma}{Lemma}
\usepackage{framed}
\usepackage{mathtools}

\begin{document}

\title{Chemical herding as a multiplicative factor for top-down manipulation of colloids}

\author{Mark N. McDonald}
\affiliation{Brigham Young University, Department of Chemical Engineering}
\author{Cameron K. Peterson}
\affiliation{Brigham Young University, Department of Electrical and Computer Engineering}
\author{Douglas R. Tree}
\affiliation{Brigham Young University, Department of Chemical Engineering}
\email{tree.doug@byu.edu}

\begin{abstract}
Colloidal particles can create reconfigurable nanomaterials, with applications such as color-changing, self-repairing, and self-regulating materials and reconfigurable drug delivery systems. However, top-down methods for manipulating colloids are limited in the scale they can control.
We consider here a new method for using chemical reactions to multiply the effects of existing top-down colloidal manipulation methods to arrange large numbers of colloids with single-particle precision, which we refer to as chemical herding. 
Using simulation-based methods, we show that if a set of chemically active colloids (herders) can be steered using external forces (i.e. electrophoretic, dielectrophoretic, magnetic, or optical forces), then a larger set of colloids (followers) that move in response to the chemical gradients produced by the herders can be steered using the control algorithms given in this paper. 
We also derive bounds that predict the maximum number of particles that can be steered in this way, and we illustrate the effectiveness of this approach using Brownian dynamics simulations. 
Based on the theoretical results and simulations, we conclude that chemical herding is a viable method for multiplying the effects of existing colloidal manipulation methods to create useful structures and materials.
\end{abstract}

\maketitle

\section{Introduction}
Colloids are ideal building blocks for the next generation of reconfigurable nanomaterials. 
Researchers have already demonstrated colloidal micromachines~\cite{Aubret2018,Aubret2021}, swarms of microrobots~\cite{Soto2022,Xie2019}, and groups of light-controlled micromotors~\cite{Palagi2019}, all of which can reconfigure their structure.
Such reconfigurable materials pave the way for advanced technologies such as color-changing materials~\cite{Liljestrom2019}, self-repairing and self-regulating materials~\cite{Fu2017,Zhang2020}, and reconfigurable drug delivery systems~\cite{ Yadav2020, Gao2014, Singh2021}. 
These reconfigurable systems can be understood using the paradigm of top-down control of bottom-up (self-assembly) processes. However, top-down (i.e. human controllable) forces tend to be limited in either the amount of local control they can apply or on the scale they can control~\cite{Isaacoff2017}. For example, direct printing can only make static arrangements that are not reconfigurable, magnetic and fluidic forces tend to act globally and can upset areas of the domain that have already been configured, and local actuators such as optical tweezers can only control a small number of particles at a time.
To address this last challenge, we propose using chemical forces in combination with existing top-down techniques to facilitate the precise, local control of larger numbers of particles. 
This approach has the potential to advance the technologies capable of developing reconfigurable colloidal materials.

Individual colloidal particles can be moved using electric fields~\cite{Armani2006,Cummins2013,Matei2019}, fluid flow~\cite{Shenoy2016, Shenoy2019, Kumar2019}, magnetic forces~\cite{Gosse2002}, optical forces~\cite{Ashkin1986,Moffitt2008}, and acoustic forces~\cite{Marzo2019,Nilsson2009,Ding2012}. 
Recent studies have also demonstrated colloids that move in response to light-controlled local chemical reactions~\cite{Aubret2018, Aubret2021,Yang2018}, and we have recently shown how individual particles can be steered using highly controllable chemical reactions in a microfluidic device~\cite{McDonald2023a}. 
Chemical reactions are an intriguing method of particle steering because natural biological systems are known to move and adapt in response to chemical signals~\cite{Kunche2016}, and because chemical reactions provide new and unexplored degrees of freedom that can be used in tandem with optical, magnetic, electric, or fluidic forces. 
For example, chemical forces may act on particles that do not have the required dielectric or magnetic properties to be manipulated directly using optical or magnetic tweezers. Even more significantly, 
we expect that chemical forces can be used to increase the number of particles that can be manipulated through other top-down methods and, for example, make relatively inexpensive electrophoretic steering techniques viable for a wider range of applications.
In this paper, we use simulation-based methods to present a new technique for using chemical reactions to multiply the effects of other single-particle colloidal manipulation methods by using a small number of chemically active particles to control a larger group of nonreactive colloids. 
We refer to this method as chemical herding. 

\begin{figure}[tbhp]
\includegraphics[width=2.50in]{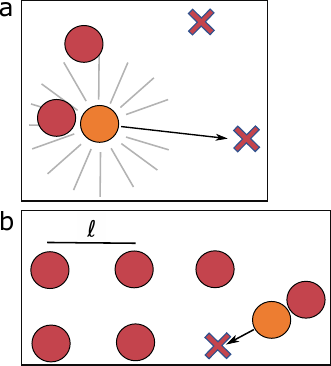}
  \caption{a) An illustration of chemical herding. We wish to move passive ``follower'' particles (red circles) to target positions (red x's), with this desired motion represented by a dotted black line. A chemically active ``herder'' particle (orange circle) creates a chemical gradient that attracts the followers through diffusiophoretic interactions. We use external forces to move the herder on a path (solid black line) that allows it to lead a follower to its target. b) Followers are made to move to pre-designated positions, separated by a distance of $\ell$. 
  }
  \label{fig-IntroHerding}
\end{figure}

Chemical herding is inspired by a technique used for unmanned aerial vehicles (UAVs) called indirect herding~\cite{Licitra2017, Licitra2018, Licitra2019}.
Indirect herding uses directly controllable agents (referred to as herders) to move passive agents (followers) to a desired location. 
While the small size of colloidal particles prevents them from being controlled using the same technologies as UAVs, the principle of indirect herding can still be applied. 
In chemical herding, the herder is a colloid that catalyzes a chemical reaction to create a solute concentration gradient, and the followers are nonreactive colloids. This situation is illustrated in Figure~\ref{fig-IntroHerding}. 
The followers (red circles) are attracted to or repelled from the herder (orange circle) by diffusiophoretic interactions with the solute (illustrated by gray lines).  
The herder is directly moved using external forces, and it is made to herd the followers to target locations (red x's) through interactions with the solute it produces. The herder is capable of moving many followers, one at a time, as illustrated by Figure~\ref{fig-IntroHerding}b. In this paper, we will move colloids onto regularly spaced, pre-defined target locations, as shown in the figure. 

\begin{figure}[hbtp] 
\includegraphics[width=3.25in]{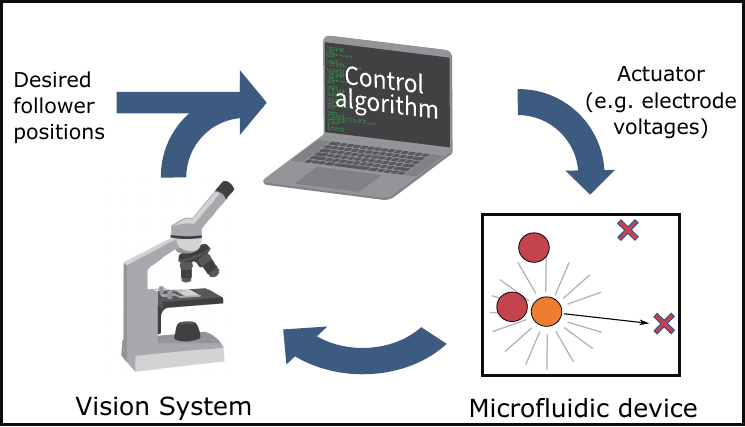}
  \caption{A diagram that explains the steps of chemical herding. 
  Six follower particles (red circles) are moved through interactions with a herder particle (orange circle) in a microfluidic device. 
  The position of each particle is measured by a vision system and supplied to a control algorithm to calculate values for the actuator that will move the herder on its calculated path.  The actuator implements a force or field that will move the herder on its desired path.  
  }
  \label{fig-Intro}
\end{figure}

We present simulations of chemical herding using Brownian Dynamics (BD) techniques. 
The BD simulations model a physical system with the following properties, illustrated in Figure~\ref{fig-Intro}: 
\begin{enumerate}
\item A vision system that measures the position of each particle in real-time, 
\item A top-down manipulation method such as optical tweezers or electrokinetic actuators that can be used to steer the herder, and 
\item A microfluidic device filled with colloidal particles in a solvent/solute system.
\end{enumerate}
Finally, in addition to the simulated system above, we add
\begin{enumerate}
  \setcounter{enumi}{3}
  \item A feedback controller to calculate the optimal values for the actuators (e.g. electrode voltages or optical trap position) using information from the vision system.
\end{enumerate}
While this paper contains only simulation results, a physical system that employs chemical herding would include the following elements: followers that are nonreactive colloids such as silica or polystyrene, a herder that is a metal colloid that catalyzes an H$_2$O$_2$ reaction, which has been seen in literature to attract other particles~\cite{Zhang2021,Singh2017,Aubret2018},
and external forces to move the herder implemented using optical tweezers, magnetic tweezers, or electrokinetic forces~\cite{Armani2006}. 
In our simulations, we have attempted to choose physical parameters that replicate physical conditions as closely as possible.

In the remainder of this paper, we will introduce our feedback controller and present simulations demonstrating the steering of colloids using chemical herding. 
In Section~\ref{sec-methods}, we will explain the simulation methods and controller. 
In Section~\ref{sec-theory}, we will use Lyapunov stability theory to derive bounds on the range of physical parameters that can be used for chemical herding. 
Then, in Section~\ref{sec-simulations} we will present the results of the simulations, including steering many particles with a single herder and using multiple herders working together. 
We will end with our conclusions in Section~\ref{sec-conclusions}.

\section{Methods} \label{sec-methods}
In this section, we will first briefly describe our Brownian Dynamics (BD) simulations, including the methods for calculating the chemical concentration profile for diffusiophoresis and implementing interparticle interactions. 
Afterward, we will explain the controller used for chemical herding. 
Then, we will introduce concepts from Lyapunov stability theory that we will use to derive the maximum number of particles that can be steered using a single herder. 

For the remainder of this paper, we will use the following notational conventions. We will
represent vectors in bold, with the norm of the vector being non-bold with a tilde, meaning the vector $\bm{r}$ has a norm $\tilde{r}$. We will use the subscript $i$ as the index for an arbitrary follower particle, and the subscript $c$ for the follower that is currently being chased, or herded, to its target. Additionally, the subscript $f$ refers to followers and $h$ refers to herders. 

\subsection{Brownian dynamics simulation methods} \label{sec-BDeqs}
Our simulations consider $n_{h}$ reactive colloidal particles (herders) with radius $R_{h}$ and $n_{f}$ nonreactive colloidal particles (followers) with radius $R_{f}$. 
The motion of these particles is determined using BD simulations.
We limit the motion of the colloidal particles to the $z=0$ plane. 
While the colloids move in quasi-2D, we will use a chemical concentration field that diffuses in 3D space to be physically realistic.

The Brownian Dynamics equation of motion for a colloidal particle is given by~\cite{Dorfman2014, Ottinger1996}
\begin{equation}
\label{eq-brownianDynamics1}
    \frac{d\bm{r}_i}{dt} =  \frac{1}{\gamma_{i}} \bm{F}_{i} + \sqrt{2 D_{i}} \bm{\xi}_{i},
\end{equation}
where $\bm{r}_i$ is the position vector of particle $i$, $\gamma_{i}$ is the friction coefficient and $D_{i}$ is the diffusion coefficient of the particle, $\bm{\xi}_{i}$ is a Gaussian white noise term, and $\bm{F}_{i}$ is the sum of non-Brownian forces acting on the particle.
We consider forces from diffusiophoresis, interparticle interactions, and externally applied forces (such as electrophoresis or optical tweezers). Diffusiophoresis accounts for the interactions between the herder and the followers. 
Diffusiophoresis in a non-ionic solute~\cite{Anderson1989} can be modeled using  
\begin{equation}
\bm{F}_{\mathrm{diff},i} = \gamma_{i} \mu_{i} \nabla C_s(\bm{r}_i)
\end{equation}
where $\mu_{i}$ is the diffusiophoretic mobility and $C_{s}$ is the concentration of the solute. 
Externally applied forces depend on the method of actuating the herder, and will be designated as $\bm{F}_{\mathrm{ext},i}$.
Interactions between particles are modeled using the Heyes--Melrose algorithm~\cite{Heyes1993}, with the interaction force given by 
\begin{equation}
    \bm{F}_{\mathrm{int},i} = \begin{dcases}
        \frac{-\gamma_i}{\Delta t_{\mathrm{sim}}} \sum_j \kappa ( R_{i} + R_{j}-\tilde{d}_{ij}) \frac{\bm{d}_{ij}}{\tilde{d}_{ij}} & \tilde{d}_{ij} < R_{i} + R_{j} \\
        0 & \mathrm{otherwise},
    \end{dcases}
\end{equation}
where $\bm{d}_{ij}$ is the vector from particle $i$ to particle $j$, $\Delta t_{\mathrm{sim}}$ is the timestep of the simulation, $R_i$ is the radius of particle $i$, $R_j$ is the radius of particle $j$, and $\kappa$ is a constant. 
Following Heyes and Melrose, we have chosen a value of $\kappa = 1.0$.

All other forces, including hydrodynamic flows, are not accounted for in this model.
Although colloids induce fluid motion which affect solvent concentration and create additional forces on each particle, accounting for hydrodynamics introduces significant computational expense and analytical difficulties due to the complex coupling between hydrodynamics and concentration profiles. 
Therefore, we have chosen to neglect hydrodynamic interactions at present so that we can emphasize the development and validation of our feedback controller.
We further hypothesize that the method of chemical herding described here may prove robust to the addition of hydrodynamic flows, because the feedback controller may compensate for errors introduced by their neglect.
We plan to test this assumption in future studies.

Using the above assumptions, Equation~\eqref{eq-brownianDynamics1} can be expressed as 
\begin{equation}
\label{eq-brownianDynamics0}
\frac{d\bm{r}_i}{dt} = \mu_{i} \nabla C_{s}(\bm{r}_i) +\frac{\bm{F}_{\mathrm{int},i} }{\gamma_{i}} +\frac{\bm{F}_{\mathrm{ext},i} }{\gamma_{i}} + \sqrt{2 D_{i}} \bm{\xi}_{i}.
\end{equation}
Our simulations contain two types of particles: followers and herders, and for each of these two types of particles, we can make further simplifications.

The forces acting on a follower particle are diffusiophoresis and interparticle interactions. 
We assume the followers are not impacted by the external forces from the actuator that steers the herders, i.e. optical/magnetic tweezers or electrokinetic forces, which primarily affect the herders. 
 Thus the equation of motion of the followers is
\begin{equation}
\label{eq-brownianDynamics2}
\frac{d\bm{x}_i}{dt} = \mu_{f} \nabla C_{s}(\bm{x}_i) +\frac{\bm{F}_{\mathrm{int},i} }{\gamma_{f}} + \sqrt{2 D_{f}} \bm{\xi}_{i},
\end{equation}
where $\bm{x}_{i}$ is the position vector of follower particle $i$, $\mu_{f}$ is the diffusiophoretic mobility of a follower particle in the solute, $C_{s}$ is the concentration of the solute, $\gamma_{f}$ is the friction coefficient and $D_{f}$ is the diffusion coefficient of the followers, and  $\bm{F}_{\mathrm{int},i}$ is the force of interparticle interactions felt by the particle.
 
 A herder particle experiences externally applied forces for steering, $\bm{F}_{\mathrm{ext},h}$, and interparticle interactions, $\bm{F}_{\mathrm{int},h}$.
 To simplify the following analysis, we assume a single herder.
 We also neglect diffusiophoretic forces on the herder (i.e. self-diffusiophoresis).
 Applying these simplifications to Equation~\eqref{eq-brownianDynamics0} gives the equation of motion for a herder particle
 \begin{equation}
\label{eq-brownianDynamics3}
    \frac{d\bm{y}}{dt} = \frac{\bm{F}_{\mathrm{ext},h} }{\gamma_{h}} +\frac{\bm{F}_{\mathrm{int},h} }{\gamma_{h}}  + \sqrt{2 D_{h}} \bm{\xi}_{h},
\end{equation}
where $\bm{y}$ is the position vector of the herder with diffusion coefficient $D_{h}$, the friction coefficient is $\gamma_{h}$, and $\bm{\xi}_{h}$ is a Gaussian white noise term. 

Assuming the reaction only occurs on the surface  of a spherical herder, the solute concentration $C_{s}$ is determined by solving the reaction-diffusion equation,
\begin{equation} \label{eq-reactiondiffusion}
    \frac{\partial C_{s}(\bm{X},t)}{\partial t} = D_s \nabla^2 C_{s}(\bm{X},t) + g_{h} \delta(\bm{X}-\bm{y}),
\end{equation}
where $\bm{X}$ is the spatial coordinate, $D_s$ is the solute diffusion coefficient, $\delta$ is the Dirac delta function, and $g_{h}$ is the rate of solute production on the surface of the herder.

We find the concentration profile around a herder by applying a pseudo-steady state approximation ($\partial C_s/\partial t \approx 0$) to Equation~\eqref{eq-reactiondiffusion}. 
If the boundary condition for concentration is $C_{\infty}$ at a distance of $||\bm{X}-\bm{y}|| \rightarrow \infty$, then
Equation~\eqref{eq-reactiondiffusion}, in the steady state limit,  has a solution of~\cite{Haberman2004}
\begin{equation}
\label{eq-concentration}
    C_s(\bm{X}) \approx  \frac{g_{h} } {4\pi D_s ||\bm{X}-\bm{y}||} + C_{\infty},
\end{equation}
with gradient
\begin{equation}
\label{eq-concentrationgradient}
    \nabla C_s(\bm{X}) \approx  \frac{-g_{h} (\bm{X}-\bm{y}) } {4\pi D_s ||\bm{X}-\bm{y}||^3}.
\end{equation}
These expressions are valid in the far field limit, and it has been shown that such a pseudo-steady state approximation gives a useful limit for modeling attractive phoretic interactions~\cite{Liebchen2017, Liebchen2019}. 
We give further justification of the pseudo-steady state approximation in the Supporting Information. 

We now have all that we need to create our BD simulations. 
The dynamics of a follower are given by Equation~\eqref{eq-brownianDynamics2}, using the $\nabla C_s$ from 
Equation~\eqref{eq-concentrationgradient}. 
The dynamics of a herder are given by Equation~\eqref{eq-brownianDynamics3}, with $\bm{F}_{\mathrm{ext},h}$ coming from the controller we will develop in Section~\ref{sec-control}.

\subsection{Brownian dynamics model parameters} \label{sec-params}

We model the herder as a platinum-coated colloidal particle that catalyzes the reaction of H$_2$O$_2$. Table~\ref{table-table1} presents the physical parameters used in our simulations. 
In the following paragraphs, we will provide a detailed explanation of the selection of values for each parameter. These values serve as a base case, and they will be varied in our later analysis. 

\begin{table}[htbp]
\caption{Physical parameters used in our BD simulations.}
\begin{tabular}{c c c}
\hline
Symbol & Explanation & Value \\
\hline
$J_{h}$ & Reaction flux of herder & 0.02 mol/m$^2$ s \\
$D_s$ & Diffusion coefficient of solute & $2.01 \times 10^{-9}$ m$^2$/s \\
$\mu_{f}$ & Diffusiophoretic mobility & $2.0 \times 10^{-10}$ m$^2$/M s \\
T & Temperature & 298K \\
$\eta$ & Solvent viscosity & 89 cP \\
$v_{\mathrm{max}}$ & Maximum speed of herder & 5 $\mu$m/s \\
$R_{f}$ & Radius of followers & 4$\mu$m \\
$R_{h}$ & Radius of herders & 4$\mu$m \\
\hline
\end{tabular} 
\label{table-table1}
\end{table}

The herder produces solute at a constant rate $g_{h}$.  For a spherical herder, we can write $g_{h} = 4 \pi r_{h}^2 J_{h}$, where $J_{h}$ is the flux of solute from the surface of the herder. 
We set $J_{h}$ based on the reaction rate of hydrogen peroxide to a platinum catalyst, which we have taken as a prototype reaction for chemical herding. 
The reaction surface flux of platinum in 10\% H$_2$O$_2$ is approximately 0.02 mol/m$^2$s~\cite{Zhou2021}. 
This gives $g_{h} = 0.02 \times 4\pi R_{h}^2 \approx 0.25R_{h}^2 $ mol/s.

Experimental observations indicate that particles that catalyze H$_2$O$_2$ create gradients that tend to attract other particles~\cite{Zhang2021,Singh2017,Aubret2018}. Thus, we use a positive value for the diffusiophoretic mobility $\mu_{f}$ of the follower particles, with the magnitude of $\mu_{f}$ based on a typical non-ionic solute,  as predicted by Anderson~\cite{Anderson1989}.
To simplify our simulations and analysis, we modeled the reaction as a single solute species, which allowed us to use a single value for the diffusiophoretic mobility.
We also set the diffusion coefficient of the solute, $D_{s}$, as the diffusion coefficient of O$_2$ in water. 

The values for the diffusion coefficient of the followers, $D_{f}$, and for the herders, $D_{h}$, were determined using the Stokes-Einstein relations 
\begin{align}
D_{f} = \frac{k_{b}T}{6 \pi \eta R_{f}} \\
D_{h} = \frac{k_{b}T}{6 \pi \eta R_{h}},
\end{align}
where $k_{b}$ is Boltzmann's constant, $T$ is the temperature, $\eta$ is the viscosity of the solvent, and $R_{f}$ and $R_{h}$ are the radii of the followers and herders, respectively. 
For our simulations, we used a temperature of 298K and the solvent viscosity of water. 
Both $R_h$ and $R_f$ were set to 4$\mu$m, which is a reasonable size for colloidal particles that can be observed using an optical microscope. 

Finally, the upper limit on the velocity of the herder is determined by the maximum speed a physical actuator can achieve,
 and is set to a fixed value of $v_{\mathrm{max}}$. 
We set $v_{\mathrm{max}}$ to 5~$\mu$m/s to ensure that the top-down forces applied to the system are not exceeded, which is a speed easily achievable by using electrode voltages to steer colloids, as reported by Armani~\cite{Armani2006}.

\subsection{Switched systems control} \label{sec-control}

In this section, we derive the control law that will be used to move the herder.
For evenly spaced target positions with a spacing of $\ell$, as shown in Figure~\ref{fig-IntroHerding}b, we wish to create a controller that will move one particle onto each target position. We will derive the algorithms to do this in the following paragraphs. 

\begin{figure*}[hbtp] 
\includegraphics[width=6.5in]{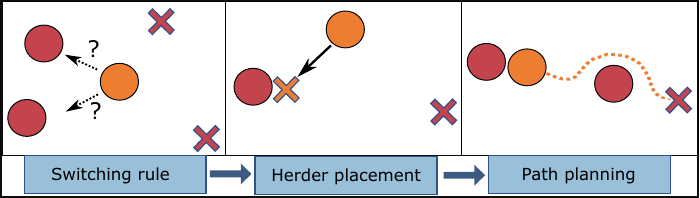}
  \caption{A schematic diagram showing the four steps in our chemical herding controller. The first panel shows our switching rule, which dictates that the herder should chase the particles one at a time, beginning with the particle farthest away from its target and then moving to the next farthest. The second panel shows the optimal placement of the herder (orange x) which will move the selected follower towards its target. The third panel shows how a GVF is used to plan the path of the herder to avoid obstacles while guiding the selected follower to its target. }
  \label{fig-HerdingExplanation}
\end{figure*}

In deriving our control law, we assume that the herder can be moved much faster using external forces than the speed at which the followers can move via diffusiophoresis. 
This allows us to decouple the path planning of the herder from the calculation of where we want the herder to be in relation to the followers. With this assumption, the herding problem can be divided into three parts: 
\begin{enumerate}
    \item setting a switching strategy for choosing which follower the herder will chase,
    \item calculating the optimal placement of the herder, and
    \item planning the path for the herder so that it avoids collisions with other particles.
\end{enumerate}
These steps are illustrated in Figure~\ref{fig-HerdingExplanation}.

In the first step, we set a switching strategy for the herder to select which follower to chase~\cite{Licitra2019}. 
We use the word ``chase'' following the terminology of Licitra et al., even though in the present work, the nature of the attractive interactions means the herder leads the follower particles rather than pursuing from behind. We will use the subscript ``$c$'' to refer to a follower being chased by the herder. 

We chose a simple and intuitive switching strategy: 
\begin{itemize}
\item Match each follower to a target using the Python algorithm \texttt{linear\_sum\_assignment} to minimize the distance to the assignment. 
\item Select the follower farthest from its target. 
\item Herd the selected particle until it is within precision $d_{\mathrm{prec}}$ of its target position.
\item Repeat until all particles are within tolerance $d_{\mathrm{tol}}$ if their target position. 
\end{itemize}
It is necessary to have two different length scales $d_{\mathrm{prec}}$ and $d_{\mathrm{tol}}$ because each deals with Brownian motion on a different time scale; $d_{\mathrm{prec}}$ considers the Brownian motion of a single time step, while $d_{\mathrm{tol}}$ must be large enough to consider Brownian motion over the period of time the herder takes to visit each follower particle. 
Values of $d_{\mathrm{prec}}$ and $d_{\mathrm{tol}}$ are given in Table~\ref{table-table2}, along with the length scale of the target pattern, $\ell$. These values will be discussed further in Section~\ref{sec-simulations}.

\begin{table}[htbp]
\caption{Length scales used in the chemical herding controller.}
\begin{tabular}{c c c}
\hline
Symbol & Explanation & Value/expression \\
\hline
$d_{\mathrm{prec}}$ & Controller precision & 1$\mu$m \\
$d_{\mathrm{tol}}$ & Controller end tolerance & 9$\mu$m \\
$\ell$ & Target spacing & 30$\mu$m \\
\hline
\end{tabular} 
\label{table-table2}
\end{table}

\begin{figure}[hbtp] 
\includegraphics[width=2.25in]{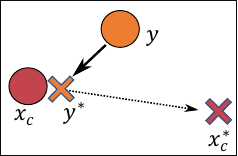}
  \caption{A schematic of a herder (orange circle) moving to its optimal placement (orange x), next to a follower (red circle). Also displayed is the target associated with that follower (red x). 
  }
  \label{fig-AlgorithmExplanation}
\end{figure}

In the second step, we calculate the optimal placement of the herder that will move the selected follower towards its target. 
 Figure~\ref{fig-AlgorithmExplanation} depicts this step graphically. 
We define ${\bm{e}_{c} = \bm{x}_c-\bm{x}_{c}^{*}}$ as the difference between the position of the currently herded follower (red circle in the figure) and its target position (red x), where the subscript $c$ refers to the follower that is being chased by the herder. 
We wish to ensure $\bm{e}_{c}$ approaches zero as quickly as possible, and we do this by placing the herder on a location $\bm{y}^{*}$ (orange x in the figure). 
Since the herder attracts the follower, placing $\bm{y}^{*}$ on the line with direction $\bm{e}_{c}/\tilde{e}_c$ leading from the particle to its target (dotted line in the figure), will attract the follower to location $\bm{x}_{c}^{*}$. 
We place the herder as close to the follower as feasible to decrease this error as quickly as possible. 
Due to hard sphere interactions, the closest that the herder can approach the follower is $R_{f} + R_{h}$, where $R_{f}$ is the radius of the follower and $R_{h}$ is the radius of the herder. 
To give the follower enough space to move, we add in a constant $0.2R_{f}$, for a total distance of $R_{fh}=R_{f}+R_h+0.2R_{f}$.
Therefore, the ideal trajectory of the herder is given by
\begin{equation}
\label{eq-herderposition}
   \bm{y}^{*}(t) =  \bm{x}_{c}(t) -\frac{\bm{e}_{c}(t)}{\tilde{e}_{c}(t)}R_{fh}. 
\end{equation}

Equation~\eqref{eq-herderposition} gives the trajectory of a herder that will move the selected follower towards its target as quickly as possible. 
However, to produce this trajectory from any given initial position would require that the herder moves arbitrarily quickly.
To relax this assumption, we instead consider Equation~\eqref{eq-herderposition} as a relationship that lets us find the optimal placement of the herder as a goal for the herder to move towards during any discrete timestep. 
The actual trajectory of the herder will be given in the following paragraphs.

For the third step, once we have determined the optimal placement for the herder, we must compute a trajectory that does not disrupt other particles or attempt to pass through the particle being herded. 
A straight-line trajectory is not suitable for this purpose. 
Instead, we adopt a path-planning approach commonly used in UAV navigation called a gradient vector field (GVF).
The GVF is a function $\bm{V}_{g}(\bm{y},\bm{y}^{*},\{\bm{x}_{i}\})$ that produces a direction for the herder to move at each time step. 
We treat the followers as obstacles that the herder must avoid and use a modified version of the GVF proposed by Wilhelm and Clem~\cite{Wilhelm2019} to calculate the direction $\bm{V}_g/\tilde{V}_g$ that will move the herder to $\bm{y}^{*}$ while avoiding collisions with the follower particles. The function $\bm{V}_g$ is defined in the Supporting Information. 
The resulting external force applied to the herder dynamics in Equation~\eqref{eq-brownianDynamics3} is set to be
\begin{equation}
\label{eq-GVFveloc}
    \bm{F}_{\mathrm{ext},h} = \gamma_{h} \frac{\bm{V}_{g}}{\tilde{V}_{g}} \min \left(v_{\mathrm{max}},\frac{\tilde{e}_{h}}{\Delta t_{\mathrm{control}} }\right),
\end{equation}
where $\tilde{e}_{h} = ||\bm{y} - \bm{y}^{*}||$ is the distance from the herder to its target position and $\Delta t_{\mathrm{control}}$ is the timestep of the controller. 

Finally, to practically implement chemical herding requires a control law to produce the $\bm{F}_{\mathrm{ext},h}$ calculated by Equation~\eqref{eq-GVFveloc} in a physical system. 
For example, electrokinetically steering the particles would need a least squares minimization algorithm to set the electrode voltages~\cite{Chaudhary2006,Armani2006}.  
However, the simulations presented in this paper do not include the specific controller and are instead independent of the method used to steer the herder, which allows our conclusions to be more generally applicable.

\section{Lyapunov stability limits on herding} \label{sec-theory}

In this section, we will introduce the concept of Lyapunov stability, then explain how to use it to derive a limit on the maximum number of particles that can be steered using a single herder. 
Then, we will show that Brownian motion creates another limit on the number of particles that can be steered, and analyze the sets of physical parameters where chemical herding is possible. 

\subsection{Lyapunov theory background}

In addition to the simulations and controller described above, we have also used Lyapunov stability analysis to derive a limit on the number of particles that can be steered.
Lyapunov theory provides a rigorous framework to ensure the stability of dynamical systems, which, in the context of chemical herding, means that followers will converge to and stay at their target locations. In the following paragraphs, we will give a brief introduction to Lyapunov stability theory. 

To illustrate Lyapunov stability, imagine we wish to confirm that some variable $x$ tends to zero. 
To facilitate this, we introduce a positive definite function $V(x)$ which attains its minimum when $x=0$. 
The crux of Lyapunov analysis then lies in demonstrating that this function, $V(x)$, consistently diminishes as time progresses, i.e., its derivative $\dot{V}(x)$ is negative definite. 
If this is achieved, it follows that $x$ evolves towards zero over time, a phenomenon termed asymptotic stability. 

For a more conceptual understanding of Lyapunov stability, one can draw a parallel between Lyapunov's stability analysis and examining the temporal evolution of a system's free energy landscape. 
If the free energy—conceptually similar to $V(x)$—diminishes with the system's evolution, it signifies that the system is inching closer to its equilibrium state, represented by $x=0$, meaning that the system is asymptotically stable. 

In this paper, we will also use the concept of exponential stability. 
Exponential stability is a stronger form of stability ensuring not just that $x$ approaches zero, but that it does so at an exponential rate, or $x \leq x(0) \exp(-\lambda t)$ for some constant $\lambda$.
Exponential stability offers robust assurances about the system's behavior, especially over short time horizons.
More information on these topics can be found in Khalil~\cite{Khalil2002}.

\subsection{Stability analysis}

We will now derive the maximum number of particles that can be steered with a single herder using Lyapunov stability analysis to find conditions for which the switched system is stable, i.e., the conditions for which the distance between the followers and their targets approaches zero as time increases. 
As shown by Licitra et al.~\cite{Licitra2018}, the stability criterion is a function of the number of followers, meaning that we can use our stability result to find the maximum number of particles that can be herded. 
To aid the reader in understanding the following analysis, we introduce the following terminology.  
By \textit{desired attraction}, we mean the force of attraction between the herder and the follower currently being chased. 
By \textit{unwanted attraction}, we mean the force of attraction between the herder and an unchased follower, which moves the follower away from its target position. 
The stability analysis will show that the relative strengths of these two forces limit the number of particles that can be steered. 

For the following analysis, we will look at the dynamics of the follower particles, which are given by Equation~\eqref{eq-brownianDynamics2}, with the concentration gradient defined in Equation~\eqref{eq-concentrationgradient}.
We will neglect Brownian motion and interparticle interactions to simplify the analysis. 
Such particles may be referred to as ``phantom'' because of the lack of interactions and ``non-Brownian'' due to the lack of Brownian motion.
Finally, we substitute Equation~\eqref{eq-concentrationgradient} into Equation~\eqref{eq-brownianDynamics2} to obtain 
\begin{equation}
\label{eq-diffusiophoreticvelocity}
\frac{d\bm{x}_i}{dt} \approx \frac{-k_{\mathrm{diff}}(\bm{x}_i-\bm{y})}{||\bm{x}_i-\bm{y}||^3} 
\end{equation} 
where 
\begin{equation}
\label{eq-kdiff}
    k_{\mathrm{diff}} = \frac{g_{h}\mu_{f}}{4\pi D_{s}} \textrm{.}
\end{equation}

Using the dynamics in Equation~\eqref{eq-diffusiophoreticvelocity}, we will prove the stability of the switched system in three parts:
First, we prove the chased particle converges exponentially to its target due to the desired attraction to the herder. 
Afterward, we show that the unchased particles remain within an exponentially bounded area around their targets, despite the unwanted attraction to the herder. 
Then, we use a theorem of switched system analysis~\cite{Yang2014} and relate the two exponential functions to show that the entire system is stable. 
Finally, we rearrange the stability criterion we derived to write a function for the maximum number of particles that can be steered. 

\subsubsection{Convergence of chased particle}
First, we will use Lyapunov stability theory to show that the currently chased follower particle will converge exponentially to its target. We consider a single follower particle $i$, with the distance from its target defined as ${\bm{e}_{i} = \bm{x}_{i}-\bm{x}_{i}^{*}}$. 
We assume that the particle is phantom and non-Brownian, so its dynamics can be described by Equation~\eqref{eq-diffusiophoreticvelocity}.

\begin{lemma}
\label{lemma-one}
Assume that the trajectory of a herder $\bm{y}(t)$ follows its optimal trajectory $\bm{y}^{*}(t)$ given by Equation~\eqref{eq-herderposition}, and that a follower with dynamics given by Equation~\eqref{eq-diffusiophoreticvelocity} is currently being herded. 
Then that follower will converge exponentially to its target with an exponential bound of 
\begin{equation}
\label{eq-convergence}
    \tilde{e}_{i}(t) \leq \tilde{e}_{i}(0)e^{-\lambda_{s} t/2},
\end{equation}
where $\lambda_{s}$ is a positive constant.
\end{lemma}

\begin{proof}
Define the Lyapunov function
\begin{equation}
\label{eq-lyap1}
    V^{s}_{i} = \frac{1}{2}\bm{e}^T_i \bm{e}_i ,
\end{equation}
with time derivative 
\begin{equation}
\label{eq-lyapintermediate}
    \dot{V}^{s}_{i} =  \bm{e}^T_{i}\dot{\bm{e}}_i.
\end{equation}
Applying Equation~\eqref{eq-diffusiophoreticvelocity}, where we recall that $\dot{\bm{e}}_{i} = d\bm{x}_i/dt$, gives
\begin{equation}
\label{eq-lyapintermediate2}
    \dot{V}^{s}_{i} =  \frac{-k_{\mathrm{diff}}}{ ||\bm{x}_i-\bm{y}||^3}\bm{e}_{i}^{T}(\bm{x}_i-\bm{y}).
\end{equation}
Substituting Equation~\eqref{eq-herderposition} into Equation~\eqref{eq-lyapintermediate2}, where particle $i$ is the currently chased particle $c$, and assuming ${\bm{y} = \bm{y}^{*}}$ yields
\begin{equation}
\label{eq-lyapintermediate3}
    \dot{V}^{s}_{i} = - \frac{k_{\mathrm{diff}}\tilde{e}_i}{R_{fh}^2 }. 
\end{equation}
Since Equation~\eqref{eq-lyapintermediate3} is negative definite, Lyapunov theory guarantees that $\bm{e}_{i}$ will decrease over time~\cite{Khalil2002}. 
This implies that
$\tilde{e}_{i}(t) \leq \tilde{e}_{i}(0)$.
Then, if we multiply the right hand side of Equation~\eqref{eq-lyapintermediate3} by $\tilde{e}_{i}(t) / \tilde{e}_{i}(0)$, we can use Equation~\ref{eq-lyap1} to write
\begin{equation}
    \dot{V}^{s}_{i} \leq -\frac{k_{\mathrm{diff}}}{R_{fh}^2} \frac{\tilde{e}_{i}^2}{\tilde{e}_{i}(0)} = -\lambda_{s}V^{s}_{i},
\end{equation}
where 
\begin{equation}
\lambda_{s} = \frac{ 2 k_{\mathrm{diff}} }{ R_{fh}^2 \tilde{e}_{i}(0)} \textrm{.} 
\end{equation}
We can then integrate to produce
\begin{equation}
    V^{s}_{i} \leq V^{s}_{i}(0) e^{-\lambda_{s}t}.
\end{equation}
Substituting in Equation~\eqref{eq-lyap1} and solving for $\bm{e}_{i}$ yields Equation~\eqref{eq-convergence}.
\end{proof}

\subsubsection{Divergence of unchased particles}
We will now show that a follower particle that is not being chased by the herder will stay within an exponentially bounded region around its target.
In other words, we write a function that gives a bound for how far a follower can wander after it has been herded.
A particle $i$ may move away from its target because, as the herder chases a different particle, particle $i$ still feels a diffusiophoretic attraction to the herder. The distance a particle can move due to this unwanted attraction to the herder is bounded as described by the following lemma. 

\begin{lemma}
\label{lemma-two}
Assume that the trajectory of a herder $\bm{y}(t)$ follows its optimal trajectory $\bm{y}^{*}(t)$ given by Equation~\eqref{eq-herderposition}, and that a follower with dynamics given by Equation~\eqref{eq-diffusiophoreticvelocity} is not currently being chased. Then that follower will remain within an exponentially bounded area around its target with an exponential bound of 
\begin{equation}
\label{eq-divergence}
    \tilde{e}_{i}(t) \leq \tilde{e}_{i}(0)e^{\lambda_{u} t/2},
\end{equation}
where $\lambda_{u}$ is a positive constant.
\end{lemma}

\begin{proof}
Consider the Lyapunov function
\begin{equation}
\label{eq-lyap2}
    V^{u}_{i} = \frac{1}{2}\bm{e}^T_{i}\bm{e}_{i}
\end{equation}
with time derivative
\begin{equation}
\label{eq-lyap2deriv}
    \dot{V}^{u}_{i} = \bm{e}^T_{i}\dot{\bm{e}}_{i}.
\end{equation}
Substituting Equation~\eqref{eq-diffusiophoreticvelocity} into Equation~\eqref{eq-lyap2deriv} as before, we get
\begin{equation}
    \dot{V}^{u}_{i} = \frac{-k_{\mathrm{diff}}}{\tilde{d}_{ih}^3} \bm{e}^T_{i}\bm{d}_{ih},
\end{equation}
where $\bm{d}_{ih}=\bm{x}_i-\bm{y}$ 
is the vector from the herder to particle $i$. 
The inner product can also be written as $\bm{e}_i^T\bm{d}_i = \tilde{e}_{i} \tilde{d}_{ih}\cos(\theta)$, where $\theta$ is the angle between the two vectors. 
The function $\dot{V}^u_i$ is maximized when $\theta=180^{\circ}$, which gives
\begin{equation}
\label{eq-lyap2intermediate}
    \dot{V}^{u}_{i} \leq \frac{k_{\mathrm{diff}}}{\tilde{d}_{ih}^2 }\tilde{e}_i.
\end{equation}

When $\theta=180^{\circ}$, $\tilde{e}_{i}$ increases with time, so we will have
$\tilde{e}_{i}(t) \geq \tilde{e}_{i}(0)$. 
Consequently, the inequality is preserved if we multiply by $\tilde{e}_{i}(t) / \tilde{e}_{i}(0)$. 
We also replace $\tilde{d}_{ih}$ with the minimum value it attains while it is not being chased by the herder, making it constant and giving us
\begin{equation}
    \dot{V}^{u}_{i} \leq \frac{k_{\mathrm{diff}}}{\min_t(\tilde{d}_{ih})^2}\frac{\tilde{e}_{i}^2}{\tilde{e}_{i}(0)} = \lambda_{u}V^{u}_{i},
\end{equation}
where
\begin{equation}
    \lambda_{u} = \frac{ 2k_{\mathrm{diff}} }{ \min_t(\tilde{d}_{ih})^2\tilde{e}_{i}(0)} \textrm{.} 
\end{equation}
We can then integrate to produce
\begin{equation}
    V^{u}_{i} \leq V^{u}_{i}(0) e^{\lambda_{u}t}.
\end{equation}
Substituting in Equation~\eqref{eq-lyap2} and solving for $\tilde{e}_i$, we see that
$\tilde{e}_i$ is exponentially bounded by 
Equation~\eqref{eq-divergence}.
\end{proof}

\subsubsection{Switched systems analysis}
We will now show that the entire switched system is stable. 
To do so, we make use of a theorem from Yang et al.~\cite[Theorem 3.1]{Yang2014} (see also~\cite[Theorem 2]{Muller2012}.)
To paraphrase the theorem, a switched system is exponentially stable if the following conditions are met. 
\begin{enumerate}
    \item One subsystem is exponentially stable with decay constant $\lambda_{s}$ and another subsystem is exponentially bounded with growth constant $\lambda_{u}$.
    
    \item The Lyapunov functions for each subsystem satisfy $V^{s}_{i} \leq \mu V^{u}_{i}$ for some $\mu \geq 1$.

    \item If $t_{s,i}$ is the time the system is stable and $t_{u,i}$ is the time the system is unstable, then there must exist some constant $\lambda\!^* \in (0,\lambda_s)$ such that
    \begin{equation}
    \label{eq-timeproof}
        \frac{t_{s,i}}{t_{u,i}} \geq \frac{\lambda_u + \lambda\!^*}{\lambda_s - \lambda\!^*}.
    \end{equation}

    \item The average dwell time $\tau_{a}$, or the average time the switched system spends in each individual subsystem, must obey 
    \begin{equation}
    \label{eq-tau}
        \tau_{a} > \frac{\ln{\mu}}{\lambda\!^*}.
    \end{equation} 
\end{enumerate}
Applying this theorem to chemical herding, Condition (1) says that the herder must drive the follower to its target at a faster rate than the follower runs away (due to unwanted attraction to the herder) when it is left alone. 
Condition (2) is a common condition in switched system analysis that says that the Lyapunov function for the unstable system cannot be of a higher order than the stable system. 
Condition (3) requires that the time a follower is unchased must be less than a certain fraction of the total time. 
Finally, condition (4) constrains how fast the herder can switch between chasing different followers. 
We will now show that, if Equation~\eqref{eq-timeproof} holds, then these conditions are true for chemical herding. 

\begin{theorem}
\label{Theorem-one}
Assume that the trajectory of a herder $\bm{y}(t)$ is equal to the optimal trajectory $\bm{y}^{*}(t)$ given by Equation~\eqref{eq-herderposition}, followers have dynamics given by Equation~\eqref{eq-diffusiophoreticvelocity},
and there exists a constant $\lambda\!^* \in (0,\lambda_s)$ such that 
Equation~\eqref{eq-timeproof} is satisfied.
Then the chemical herding system is exponentially stable.
\end{theorem}

\begin{proof}
We have already shown in Lemmas \ref{lemma-one} and \ref{lemma-two} that
Condition (1) is satisfied
with $\lambda_{s} = 2k_{\mathrm{diff}}/R_{fh}^2 \tilde{e}_{i}(0)$ and $\lambda_{u} = 2k_{\mathrm{diff}}/\min_t(\tilde{d_i})^2\tilde{e}_{i}(0)$. 
Also, from Equations~\eqref{eq-lyap1} and \eqref{eq-lyap2}, Condition (2) is satisfied with $\mu = 1$. 
Since $\ln(1)=0$, Condition (4) is trivially satisfied. 
Then, if Equation~\eqref{eq-timeproof} holds, then Condition (3) is true, and the switched system is exponentially stable. 
\end{proof}

\subsection{Limits on the number of steerable particles}

We can now derive two limits on the number of particles $n_{f}$ that can be steered. 
First, using Theorem \ref{Theorem-one}, we can derive a limit on the number of particles that can be steered by finding the conditions where Equation~\eqref{eq-timeproof} holds. 
This limit occurs because, as $n_f$ increases, the herder must spend more time moving particles back to their targets after unwanted attraction between the herder and unchased particles moves them away. 
Second, we will derive a bound on the number of particles by comparing the length scale of Brownian motion to the length scale $d_{\mathrm{tol}}$, or the distance from the targets the followers must be moved to before we consider the herding to be successful. 
This limit occurs because, as $n_f$ increases, the herder must spend more time correcting for the effects of Brownian motion. 
In this paper, we will treat these two limits independently of each other and claim that herding works as long as both limits are kept. 

\subsubsection{Limit from unwanted attraction to herder}

We now use Theorem \ref{Theorem-one} to derive a limit on the number of particles that can be steered. 
In Equation~\eqref{eq-timeproof}, the ratio of timescales $t_{u,i}$ and $t_{s,i}$ is a function of an arbitrary parameter $\lambda\!^*$. Since Equation~\eqref{eq-tau} is satisfied regardless of the value of $\lambda\!^*$, we are free to take $\lambda\!^*$ to be as small as possible. 
In the limit of $\lambda\!^* \rightarrow 0$, we can re-arrange Equation~\eqref{eq-timeproof} to give 
\begin{equation}
\label{eq-stabilityresultearly}
    \frac{t_{u,i} + t_{s,i} }{t_{s,i}} \leq 1 + \frac{\min_t(\tilde{d}_{ih})^2}{R_{fh}^2} \textrm{.}
\end{equation}

Next, we need a relationship for $t_{u,i} + t_{s,i}$. 
Note that,  ${t_{u,i} + t_{s,i}}$ equals the total time for the herder to travel between and herd each of the $n_f$ followers. 
We define the variable $t_{h}$ as the average time for the herder to travel between two followers. 
Then  
\begin{equation} \label{eq-ttot1}
    t_{u,i} + t_{s,i}  = n_f t_{h} + \sum_i^{n_f}t_{s,i}  \textrm{.} 
\end{equation}
The maximum number of particles for which the inequality in Equation~\eqref{eq-stabilityresultearly} holds occurs when each follower takes the same amount of time to move to its target. 
With this assumption, we drop the subscript $i$, giving
\begin{equation} \label{eq-ttot}
    t_{u} + t_{s} = n_{f}(t_{s}+t_{h}) \textrm{.} 
\end{equation}

Now, we will evaluate $t_h$ and $t_s$. 
To evaluate $t_h$, assume the herder moves at a constant speed $v_{\mathrm{max}}$ and an average distance $\ell$ every time it switches followers. 
Then the average time for the herder to travel between followers is 
\begin{equation}
\label{eq-th}
    t_{h} = \ell/v_{\mathrm{max}}.
\end{equation}
We can evaluate $t_{s}$ using the speed and distance that a chased particle moves.  We define  $\tilde{e}$ as the distance and $v_{\mathrm{chased}}$ as the speed a chased particle moves. 
From Equation~\eqref{eq-diffusiophoreticvelocity}, a particle that is being herded (at a distance of $ R_{fh}$ from the herder) will move at a speed of 
\begin{equation}
    v_{\mathrm{chased}} = k_{\mathrm{diff}}/R_{fh}^2 \textrm{.}
\end{equation}
Then the time for the chased particle to move a distance $\tilde{e}$ is 
\begin{equation} \label{eq-ts}
    t_{s} = \frac{ \tilde{e} }{ v_{\mathrm{chased}} } = \frac{ \tilde{e} R_{fh}^2 }{ k_{\mathrm{diff}} }  \textrm{.}
\end{equation}

Equation~\eqref{eq-stabilityresultearly} can now be evaluated by substituting in Equations~\eqref{eq-ttot}, \eqref{eq-th}, and \eqref{eq-ts}. Doing so and solving for $n_{f}$ yields
\begin{equation}
\label{eq-stabilityresulthalf}
    n_{f} \leq \left ( 1 + \frac{min_t(\tilde{d}_{ih})^2}{R_{fh}^2} \right ) \left ( 1+ \frac{l}{R_{fh}^2} \frac{k_{\mathrm{diff}}}{\tilde{e} v_{\mathrm{max}} } \right )^{-1} \textrm{.}
\end{equation}
This expression gives a limit for the number of particles that can be steered before the herder requires more time to correct for unwanted attraction than for moving the chased particle to its target. However, is not yet useful, because $\tilde{d}_{ih}$ and $\tilde{e}$ are both functions of time. Let us address these variables one at a time. 

First, $\min_t(\tilde{d}_{ih})$ is both time-varying and unrealistically restrictive.
In practice, $\min_t(\tilde{d}_{ih})$ may be equal to $R_{fh}$, the closest the herder can approach a follower.
This would imply that a particle could be dragged away at the same rate as it moves towards the target. Such behavior is never observed in our simulations. On the other hand, if the herder can move arbitrarily fast, then $\min_t(\tilde{d}_{ih})$ will be approximately equal to the distance between follower $i$ and follower $i+1$. We have already said that this distance approaches $\ell$. With this reasoning, and our empirical observations, we postulate that replacing $\min_t(\tilde{d}_{ih})$ with $\ell$ will produce a more useful bound. 

Second, the right-hand side of Equation~\eqref{eq-stabilityresulthalf} is at its tightest bound when $\tilde{e}$ is minimized, so we can substitute $\tilde{e}$ with its minimum value to get a function of constant parameters. The minimum value of $\tilde{e}$ is 
$d_{\mathrm{tol}}$, the tolerance at which we will stop the herding, as defined in Section~\ref{sec-control}, so we will replace $\tilde{e}$ with $d_{\mathrm{tol}}$.
After these substitutions, we have  
\begin{equation}
\label{eq-stabilityresulalso}
    n_{f} \leq \left ( 1 + \frac{\ell^2}{R_{fh}^2} \right ) \left ( 1+ \frac{\ell}{R_{fh}^2} \frac{k_{\mathrm{diff}}}{\tilde{e} v_{\mathrm{max}} } \right )^{-1} \textrm{.}
\end{equation}
Finally, we substitute in $k_{\mathrm{diff}}$ from Equation~\eqref{eq-kdiff} to illustrate the full set of parameters that affect $n_{f}$. 
The maximum number of particles that can be steered is then given by 
\begin{equation}
\label{eq-stabilityresultfull}
    n_{f} \leq \left ( 1+ \frac{\ell^2}{R_{fh}^2} \right ) \left ( 1+ \frac{g_h \mu_{f} \ell}{ 4\pi D_s d_{\mathrm{tol}}v_{\mathrm{max}} R_{fh}^2} \right )^{-1} \textrm{.}
\end{equation}

We note some interesting aspects of Equation~\eqref{eq-stabilityresultfull}.
To begin with, in the limit as $v_{\mathrm{max}} \rightarrow \infty$, the number of followers $n_f$ depends only on the ratio of the square of the distance $\ell$ between two targets and the square of the distance $R_{fh}$ between the herder and a chased particle, or 
\begin{equation}
    \label{eq-limitingnumber}
    \lim_{v_{\mathrm{max}}\rightarrow \infty}  n_f \leq 1 + (l/R_{fh})^2 \textrm{.} 
\end{equation}
This relationship can be explained by comparing the strengths of desired attraction and unwanted attraction. 
If $\ell$ is increased, then the unchased followers are allowed to remain farther from the herder, and the force of unwanted attraction is decreased, meaning more particles can be steered. 
If $R_{fh}$ is increased, then only the chased follower is farther from the herder, and the force of desired attraction is decreased relative to unwanted attraction, meaning fewer particles can be herded. 
Also, both desired and undesired attraction have a squared dependence on distance, as seen in Equation~\eqref{eq-diffusiophoreticvelocity}, which explains the squared relationship between $n_f$ and $\ell$ and $R_{fh}$.
However, we note that a number of approximations were made in deriving this result, and so information may have been lost. 
If the distance $\ell$ is not representative of the average distance between the herder and an unchased particle, then Equation~\eqref{eq-stabilityresultfull} may not be accurate. 
We will analyze this possibility further using simulations in Section~\ref{sec-simulations}. 

If the herder cannot move sufficiently fast ($v_{\mathrm{max}}$ is not infinite), then we must use the full form of Equation~\eqref{eq-stabilityresultfull}. 
If  $v_{\mathrm{max}}$ is of comparable size to the diffusiophoretic velocities of the followers, then the unwanted attraction between the herder and unchased particles will have time to act while the herder is moving betweeen followers, and this will decrease the number of particles that can be steered. 
The diffusiophoretic velocity, from Equation~\eqref{eq-diffusiophoreticvelocity}, is directly proportional to $g_h$ and $\mu_f$ and inversely proportional to $D_s$.
That is why, in Equation~\eqref{eq-stabilityresultfull}, the bound on $n_f$ gets tighter as $g_h$ and $\mu_f$ increase and as $D_s$ decreases.

It is also interesting to note that Equation~\eqref{eq-stabilityresultfull} is a function of $d_{\mathrm{tol}}$.  
This happens because $d_{\mathrm{tol}}$ is the length scale over which the desired attraction happens when the particles are being maintained near their targets.
If the time for the herder to travel between followers, with length scale $\ell$, is large compared to the time for the herder to interact with the followers, with length scale $d_{\mathrm{tol}}$, then this will decrease the number of followers that can be herded. Thus, if we want more precise placement of the particles (a smaller $d_{\mathrm{tol}}$), we cannot herd as many particles.

\subsubsection{Limit from Brownian Motion}

We now consider the effects of Brownian motion. 
In time $t$, Brownian motion will move a follower particle in 2D a root mean square distance of 
\begin{equation}
\label{eq-dbrown}
    d_{\mathrm{brown}}(t) = \sqrt{4 D_f t} \textrm{.}
\end{equation}
For herding to be achieved to within a tolerance of $d_{\mathrm{tol}}$, we must have 
$d_{\mathrm{brown}}$ be smaller than $d_{\mathrm{tol}}$ during that time period. 

The time for the herder to visit each particle (assuming each follower takes the same amount of time to move to its target) is $t_s + t_u$, given by Equation~\eqref{eq-ttot}. 
We must have
\begin{equation}
\label{eq-browniancheck}
   d_{\mathrm{tol}} \geq d_{\mathrm{brown}} \left(t_{s}+t_{u} \right) 
\end{equation}
for the herder to be able to move all particles to within $d_{\mathrm{tol}}$ before Brownian motion moves them away. 
Solving Equation~\eqref{eq-browniancheck} for $n_{f}$ by substituting in Equations~\eqref{eq-ttot}, \eqref{eq-th}, \eqref{eq-ts}, and
\eqref{eq-dbrown}
gives 
\begin{equation}
    \label{eq-browniancheck2}
    n_{f} \leq \frac{d_{\mathrm{tol}}^2}{4D_{f}} \left( \frac{4\pi D_s \tilde{e}R_{fh}^2}{g_h \mu_{f}} + \frac{l}{v_{\mathrm{max}}} \right)^{-1}.
\end{equation}
However, as before, we substitute $\tilde{e}$ with its minimum value of $d_{\mathrm{tol}}$ to get the tightest bound, giving
\begin{equation}
    \label{eq-browniancheck3}
    n_{f} \leq \frac{d_{\mathrm{tol}}^2}{4D_{f}} \left( \frac{4\pi D_s d_{\mathrm{tol}}R_{fh}^2}{g_h \mu_{f}} + \frac{l}{v_{\mathrm{max}}} \right)^{-1}.
\end{equation}

\subsubsection{Predicted Limits based on Combined Bounds}

Equations~\eqref{eq-stabilityresultfull} and \eqref{eq-browniancheck3} give two different bounds on the maximum number of particles that can be steered. Both bounds must be satisfied for steering to be viable.

\begin{figure*}[hbtp] 
\includegraphics[width=6.5in]{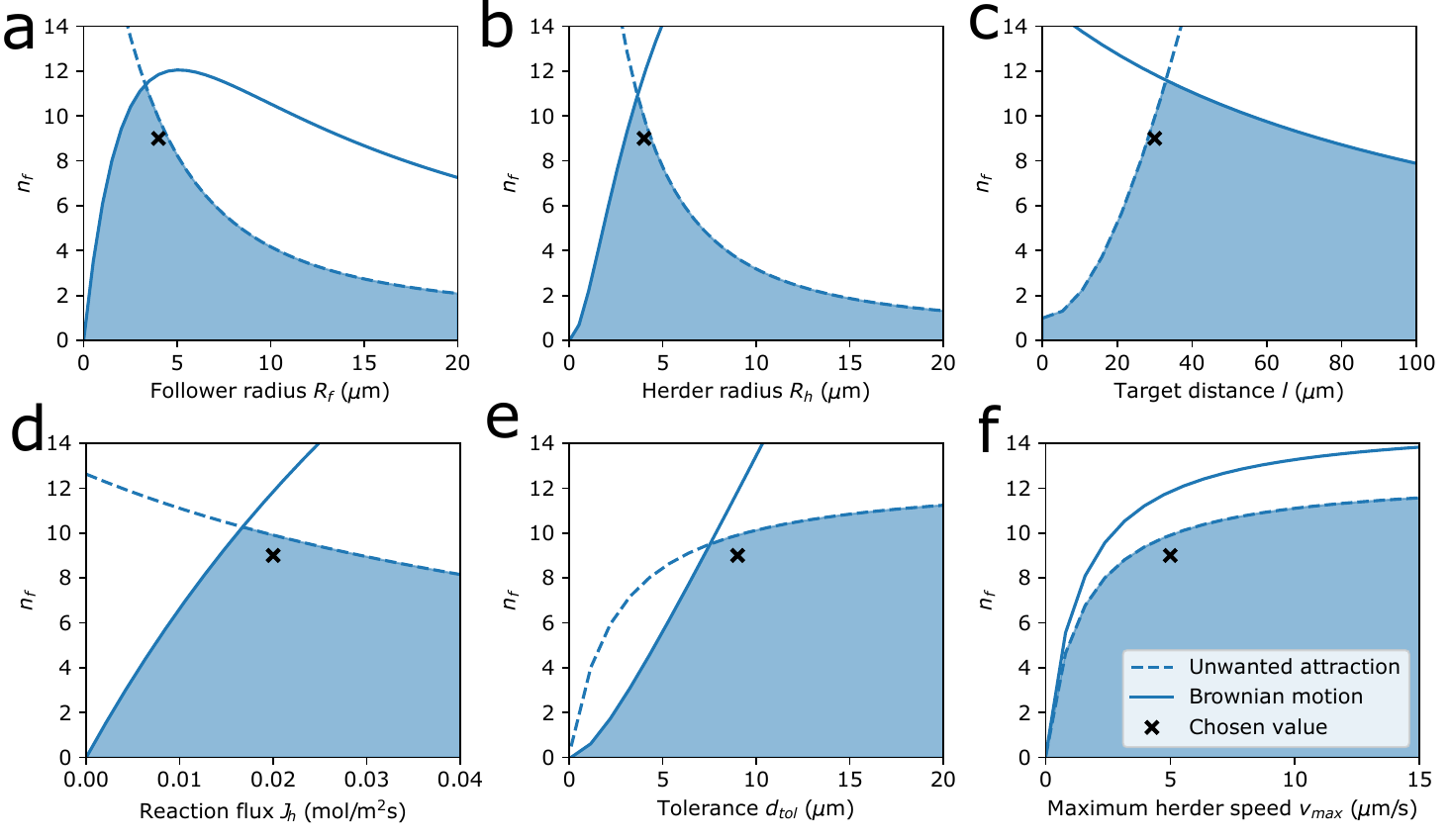}
\caption{The number of particles that can be steered as a function of six different parameters. 
In each plot, one parameter is varied, and the rest take values given in Tables~\ref{table-table1} and \ref{table-table2}.
The bound given by Equation~\eqref{eq-stabilityresultfull} is shown as a dashed line, and the bound given by Equation~\eqref{eq-browniancheck3} is shown as a solid line. 
The region in which both equations are satisfied is shaded in blue, and the choice of parameters used in the simulation below for six particles is shown as a black x.}
  \label{fig-NumberOfParticles2}
\end{figure*}

Figure~\ref{fig-NumberOfParticles2} shows the number of particles that can be steered as a function of the six  parameters $R_{f}, l, d_{\mathrm{tol}}, R_{h}, $ $J_h$, and $v_{\mathrm{max}}$. The area shaded in blue shows the set of parameters for which particle steering is viable. 
For example, Figure~\ref{fig-NumberOfParticles2}a has a maximum of $n_f \approx 11$ when $R_f$ is between 2 and 4 $\mu$m. 
Note that this plot is only valid for the specific values of the other parameters that have been selected, and different values of $\ell, d_{\mathrm{tol}}, R_{h}, $ $J_h$, or $v_{\mathrm{max}}$  may move the maximum to the left or right.

Both the follower radius $R_f$ and the herder radius $R_h$ strongly affect the number of particles that can be steered, as shown in Figure~\ref{fig-NumberOfParticles2}a and b.
Each of these plots peaks (with our chosen values of other parameters) between 2$\mu$m and 4$\mu$m. 
For smaller particle sizes, the number of particles capable of being steered falls sharply. 
At smaller sizes of $R_f$, the diffusion coefficient of the followers $D_f$ becomes large and Brownian motion moves the followers away from their targets faster than the herder can correct them.
At smaller sizes of $R_h$, the smaller reaction rate (since $g_h$ is proportional to $R_h$) makes the herding take longer, which also allows Brownian motion to dominate. 
At larger sizes of either parameter, the herder and followers are forced to be farther apart, which lowers the force of desired attraction compared to unwanted attraction and lowers the number of particles that can be steered. 
This analysis suggests that chemical herding is scale-dependent and may only be viable for a narrow range of particle sizes.

The number of particles that can be steered also depends on how close together the target positions for the particles are, represented by variable $\ell$.
As shown in Figure~\ref{fig-NumberOfParticles2}c, a target distance of less than 20$\mu$m (with our choice of other parameters) will only allow four or fewer particles to be steered, due to the limit created by unwanted attraction. 
This is because the unwanted attraction between the herder and the unchased particles gets stronger as the particles get closer together.
A target distance that is too large will also create a limit on $n_f$ due to Brownian motion.
If the targets are placed far away from each other, then the herder will take longer to move between followers and Brownian motion will have time to move followers away from their targets before the herder has a chance to correct them. 
However, with the choice of parameters we used, this upper limit is less restrictive, and up to 8 particles can still be steered at $\ell=100\mu$m. 
We wish to note that the bounds we have derived assumed a distance $\ell$ that is both the average distance between the herder and an unchased follower and also the average distance the herder must move when traveling between followers. 
Different shapes of target positions might make either of these assumptions inadequate to capture the behavior of the system. 
Thus, the shape of the target may affect the behavior of the system in ways we have not been able to fully capture with this analysis. 

The reaction flux of the solute on the surface of the herder, $J_{h}$, also affects the number of particles that can be steered, as shown in Figure~\ref{fig-NumberOfParticles2}d. We plotted $J_h$ instead of $g_h$ to separate the effects of changing the herder radius, but the two variables can be related using $g_h = 4\pi R_h^2 J_h$. As seen in Figure~\ref{fig-NumberOfParticles2}d, a reaction flux of less than about 0.005 mol/m$^2$s (for our choice of other parameters) will only allow fewer than four particles to be herded, though the bound on $n_f$ increases steadily as $J_h$ increases to about 0.02 mol/m$^2$s. Values of $J_h$ in this range are realistic for the H$_2$O$_2$ reaction we have chosen as our example. 
Increasing $J_{h}$ will increase the desired attraction between the herder and chased follower, which will increase the speed at which the herder moves a follower to its target and allow the herder to correct for Brownian motion more quickly.
But a larger $J_{h}$ will also increase the force of unwanted attraction on the unchased particles, as discussed previously. This means that if $J_h$ is too large, unwanted attraction will limit the number of particles that can be steered.

The variable $d_{\mathrm{tol}}$, the tolerance at which we conclude that the followers are close enough to their targets, also plays a role in the number of particles that can be steered, as shown in Figure~\ref{fig-NumberOfParticles2}e. In this plot, the Brownian motion curve is limiting until about 8$\mu$m, where it crosses the unwanted attraction curve. 
Since $d_{\mathrm{tol}}$ is the tolerance that we require the herding to achieve, it makes sense that a smaller tolerance will be more difficult to produce. At small values of $d_{\mathrm{tol}}$, Brownian motion moves the followers away from their targets faster than the herder can visit each follower to correct the disturbance. 
At large values of $d_{\mathrm{tol}}$ Brownian motion is not a significant factor, but unwanted attraction becomes important for reasons discussed previously, and $n_f$ approaches the asymptote of $n_f = 1+l^2/R_{fh}^2$, as given by Equation~\eqref{eq-limitingnumber}. 

Finally, the effects of the maximum speed of the herder, $v_{\mathrm{max}}$, are shown in Figure~\ref{fig-NumberOfParticles2}f. In this plot, both bounds increase monotonically, but for our choice of parameters, unwanted attraction remains the more restrictive bound. 
A larger $v_{\mathrm{max}}$ will decrease the time the herder takes to travel between followers, which will allow the herder to meet both bounds more easily. However, the effects of this increase asymptote to $n_f = 1+l^2/R_{fh}^2$, as discussed previously. 

\section{Brownian dynamics simulations} \label{sec-simulations}

\begin{figure*}[hbtp] 
\includegraphics[width=6.5in]{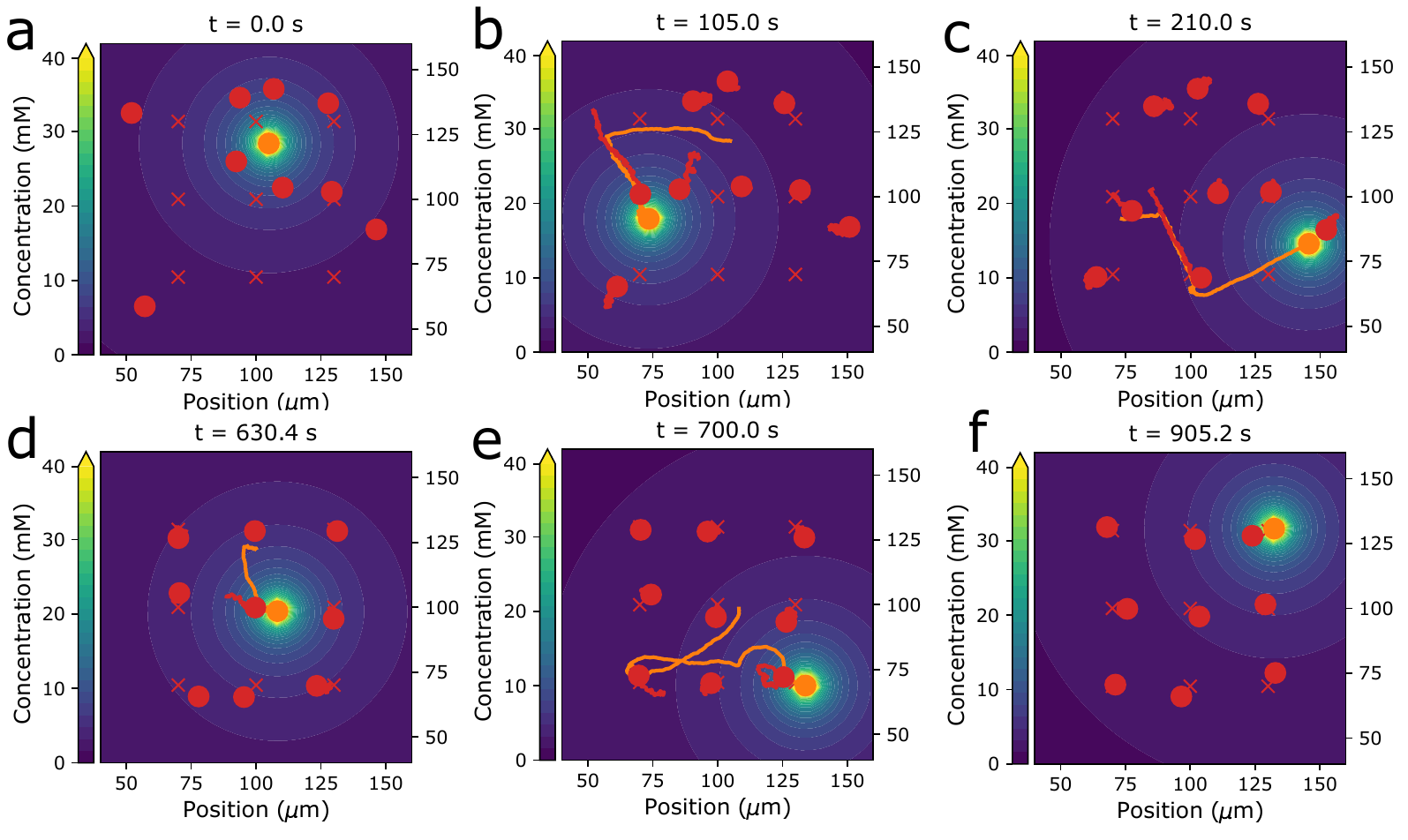}
  \caption{One herder (orange circle) is used to move six follower particles (red circles) to their associated target positions (red x's) on a lattice with a spacing $\ell=30\mu$m. Contours show the concentration produced by the herder.
Plot (a) shows the random initial condition. Plot (f) shows the particles after they have been moved to within $d_{\mathrm{tol}}$ of their target positions and maintained there for 5 minutes. Plots b-e show intermediate times. The tails (orange and red lines) behind the particles show their trajectory over different time periods, with the choice of tail length explained in the text. 
}
  \label{fig-Lattice}
\end{figure*}

Now we will demonstrate BD simulations of chemical herding. 
First, we will demonstrate a single herder steering particles from random initial positions into a lattice and into a circular formation, with parameters selected using the relationships given in the previous section.
Then we will demonstrate that it is possible to use multiple herders in tandem to move many particles at a time. 

\subsection{Single herder simulations}

We will now show simulations of chemical herding and demonstrate how the rules developed in the previous section work in practice, by looking at the time it takes to solve the herding problem under different conditions. 
We will also show that chemical herding can be used to produce different target shapes.
Then, we will look at ways to reduce the time needed for chemical herding. 

Figure~\ref{fig-Lattice} shows a chemical herding simulation where nine followers are steered from an initial random arrangement to a regular lattice with $\ell=30\mu$m spacing. The initial positions were chosen from a random uniform distribution on the portion of the domain between 50 and 150$\mu$m. 
Each particle was steered to a target using the control algorithms explained previously, and the simulations were ended after each particle was steered to within $d_{\mathrm{tol}}$ of its target, plus another five minutes to show that the herder could maintain the particles on their targets.
A full simulation is shown in Video~1.
We repeated this simulation 100 times with different initial conditions, and the example shown in Figure~\ref{fig-Lattice} is a typical result. 

The herding problem can be divided into two phases, as illustrated by Figure~\ref{fig-Lattice}: 
first, initially moving each particle to within $d_{\mathrm{tol}}$ of its target, and second, maintaining the particles near their target positions. 
Figure~\ref{fig-Lattice}a-c shows part of the initial phase, and figure~\ref{fig-Lattice}d-f shows the latter maintenance phase.

In the initial phase, the herder moves each follower from their initial position to their target position. 
Figure~\ref{fig-Lattice}a shows the initial positions of the particles. 
Figure~\ref{fig-Lattice}b shows the trajectory of each particle as the herder moves the first follower to its target. According to our switching rule, the follower farthest from its target is chased first.
Then, once the first follower reaches its target, the herder switches to chasing another follower that is now the farthest from its target at the new time.
Figure~\ref{fig-Lattice}c shows the trajectories of each particle as the herder moves a second particle to its target and begins chasing a third particle. 

Herding continues similarly until all particles are within $d_{\mathrm{tol}}$ of their targets, and then a maintenance phase begins.
Figure~\ref{fig-Lattice}d shows the positions of the particles after each particle has been moved to within $d_{\mathrm{tol}}$ of its target position. 
Brownian motion and unwanted attraction continue to affect the particles, so the herder must continue to herd the followers to maintain their positions. 
Figure~\ref{fig-Lattice}e shows how the herder uses the same switching rule and control algorithms to maintain the particles on their target positions and correct for the Brownian motion moving the particles away from their targets. 
And Figure~\ref{fig-Lattice}f shows the positions of the particles five minutes after the particles reached $d_{\mathrm{tol}}$, illustrating that the arrangement of particles can be maintained. 

\begin{figure*}[hbtp] 
\includegraphics[width=6.5in]{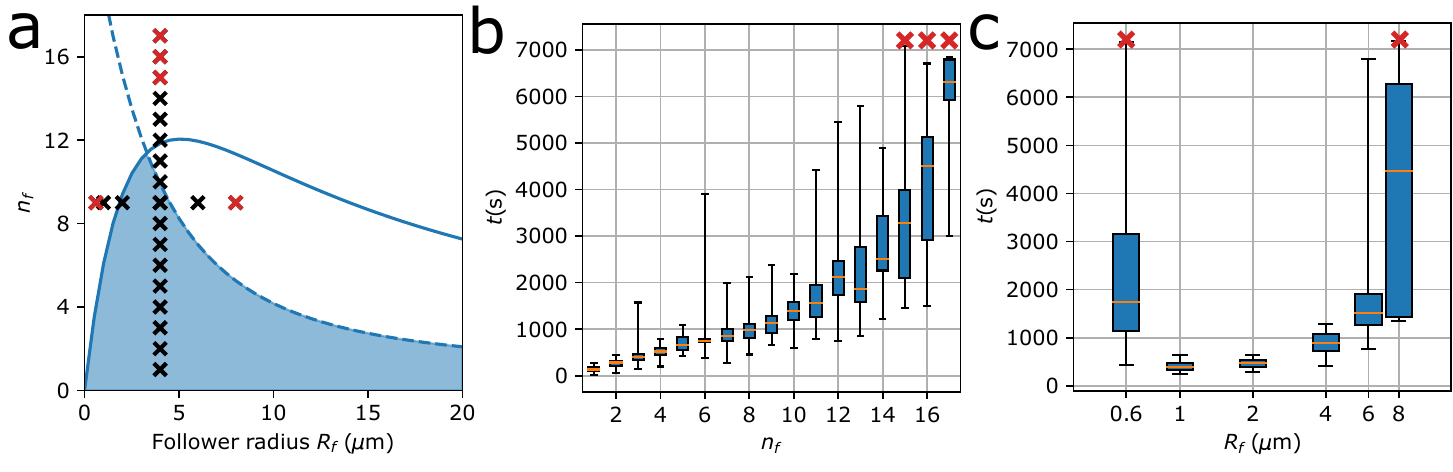} 
  \caption{Statistics of BD simulations performed over a range of parameters. 
  Plot (a) shows the parameter values for the simulations. Plot (b) shows box plots of the time to solve at each value of $n_f$. 
  Red x markers represent simulations that did not end in the two-hour time limit. 
  Plot (c) shows box plots of the time to solve at each value of $R_f$. Again, red x markers represent simulations that did not end in the two-hour time limit. 
}
  \label{fig-MonteCarlo}
\end{figure*}

Using similar simulations, we tested the theoretical predictions made in the previous section that Equations~\eqref{eq-stabilityresultfull} and \eqref{eq-browniancheck3} provide bounds for the number of particles $n_f$ that can be steered as a function of different parameters. 
To do this, we ran simulations for many values of the number of followers $n_f$ and the radius of a herder $R_f$, and tracked the time to completion.  
If the herder took longer than two hours (7200 seconds) to move all followers to their targets, we ended the simulation, reasoning that two hours would be an unrealistically long time to perform this type of experiment in a physical system. 
We performed 100 iterations for each $n_f$ between 1 and 17, and for each of six different radii, with the results shown in Figure~\ref{fig-MonteCarlo}.
Figure~\ref{fig-MonteCarlo}a shows the parameter values used in our simulations compared to the bounds predicted by Equations~\eqref{eq-stabilityresultfull} and \eqref{eq-browniancheck3} from Figure~\ref{fig-NumberOfParticles2}a.
Figure~\ref{fig-MonteCarlo}b shows the time to reach the desired configuration as a function of $n_f$, and Figure~\ref{fig-MonteCarlo}c shows the time as a function of $R_f$. 

The results in Figure~\ref{fig-MonteCarlo}a and Figure~\ref{fig-MonteCarlo}b where the number of follower particles ($n_{f}$) is varied shows that the bounds we predicted by Lyapunov stability theory are close but conservative. 
The theoretical bounds in Figure~\ref{fig-MonteCarlo}a predict that only 10 particles can be steered, but in Figure~\ref{fig-MonteCarlo}b, up to 14 particles were consistently moved to their targets within the two-hour time limit. 
For $n_f = 15$, five (out of 100) simulations did not finish, for $n_f=16$, 46 simulations did not finish, and for $n_f=17$, 65 simulations did not finish within the two-hour time limit.
Thus the bounds we derived can be interpreted as a conservative estimate of how many particles can be steered in a reasonable amount of time.

The results in Figure~\ref{fig-MonteCarlo}a and Figure~\ref{fig-MonteCarlo}c that show variation in the follower radius $R_{f}$ demonstrate that both qualitative effects predicted by theory---unwanted attraction and Brownian motion---are important bounds.
Figure~\ref{fig-MonteCarlo}c shows the time it took the herder to move nine particles onto a lattice for radii of 0.6, 1, 2, 4, 6, and 8 $\mu$m. 
The time gets very large for radii that are too small or too large. 
For both $R_f$=0.6$\mu$m and $R_f = 8 \mu$m, at least some simulations were cut off at the two-hour mark. 
For $R_f$=0.6$\mu$m, three simulations (out of 100) exceeded the two-hour limit, and for $R_f=$8 $\mu$m, 91 of the simulations exceeded the two-hour limit.
This shows that there is both a bound on how small and how large follower particles can be, as predicted by our theory.

\begin{figure*}[hbtp]
\includegraphics[width=6.5in]{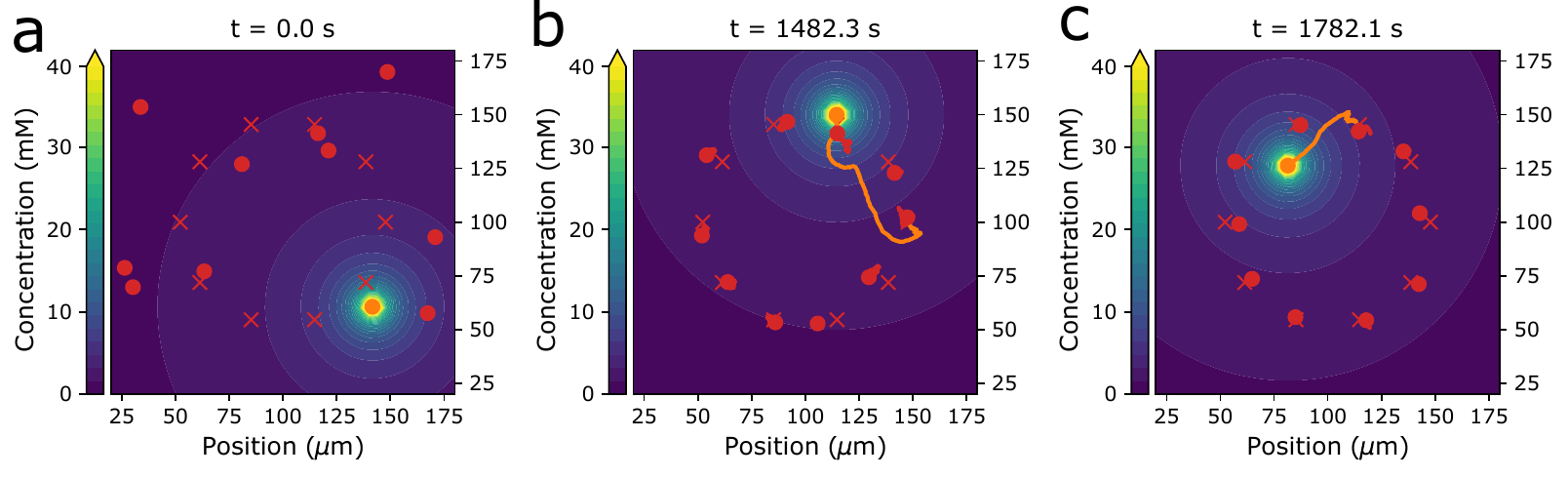} 
  \caption{One herder (orange) is used to move ten followers (red) to their target positions (red x's), arranged in a circle with a spacing between targets $\ell = 30\mu$m.
Plot (a) shows the random initial condition. Plot (b) shows the particles after the herder has moved each particle to within $d_{\mathrm{tol}}$ of their targets. Plot (c) shows after it has maintained them there for 5 minutes.
}
  \label{fig-Circle}
\end{figure*}

We also investigated patterns other than a lattice.
Figure~\ref{fig-Circle} shows a simulation in which a single herder steers 10 particles into a circular pattern. 
Particles started from initial positions taken from a random uniform distribution on the portion of the domain between 25$\mu$m and 175$\mu$m. 
The full simulation is shown in Video~2.
We again used the parameters from Table~\ref{table-table1}, which, as seen from Figure~\ref{fig-NumberOfParticles2}, allows us to steer up to 10 particles with a single herder without violating the constraints given by Equations~\eqref{eq-stabilityresultfull} and \eqref{eq-browniancheck3}.
As previously noted, these bounds are only approximate; we have observed some situations in which a greater number of particles can be steered.
The shape of the target arrangement likely impacts the actual number of particles that can be herded.
We expect that, in general, when the average distance between targets is much greater than the minimum distance $\ell$, Equations~\eqref{eq-stabilityresultfull} and \eqref{eq-browniancheck3} will not adequately predict the number of particles that can be steered.

Another important consideration in chemical herding is the time it takes the system to reach the desired configuration. 
For the simulations where ten particles were herded into a circle, the average time to converge was 1200 seconds with a standard deviation of 370 seconds (in 100 runs with random initial conditions). 
In the simulation shown in Figure~\ref{fig-Circle}, it took 1480 seconds to move all ten particles to within $d_{\mathrm{tol}}$ of their targets, which may be longer than convenient for many types of experiments. 

There are several methods that may potentially increase the speed of the chemical herding system.
First, a more efficient switching strategy could be chosen.
In Video~2, there are numerous times where the herder travels across the diameter of the circle to chase the next particle, when it would be preferable to first chase a closer particle.
A more efficient switching strategy could account for the distance the herder must travel.

Second, unwanted attraction to the herder could be leveraged to speed up chemical herding.
Unwanted attraction causes the followers to be pulled off their targets as the herder chases other particles. 
But for some target orientations, like the circle, the particles could be placed farther out from their target positions, knowing that the unwanted attraction would tend to move them towards the center.
This observation suggests that a more advanced control technique such as model predictive control could be used to plan the trajectory of the herder to use unwanted attraction as beneficially as possible.

Third, physical parameters could be selected to cause the herder and followers to move faster.
When a herder travels between followers, the maximum speed $v_{\mathrm{max}}$ it can travel is constrained by how much force (e.g., electrophoretic forces on the herder) can be applied by the actuator.
A more powerful actuator would increase $v_{\mathrm{max}}$ and thereby reduce the time for the herder to travel between followers.
More importantly, much of the time spent in chemical herding is at the slower diffusiophoretic speed of the follower particles.
The time for diffusiophoresis to move a follower a given distance is given by Equation~\eqref{eq-ts}.
Thus a larger mobility, a higher reaction rate, a smaller solute diffusion coefficient, or smaller herder and follower sizes would increase the herding speed.

Fourth, the chemistry of the physical system could affect the amount of unwanted attraction present. 
For example, a bulk reaction that consumes the solute as it diffuses away from the herder could serve to reduce the unwanted attraction between the herder and unchased followers. 
Such a bulk reaction has been referred to as ``chemical screening''.~\cite{Liebchen2019, Huang2017} 
If the solute is consumed by a bulk reaction, then that would decrease the strength of the gradient that is felt far from the herder, which could allow us to greatly relax the bound created by unwanted attraction and also save time by reducing the number of times the herder must go back and correct the positions of followers.

\subsection{Multiple herders}

\begin{figure*}[hbtp]
\includegraphics[width=6.5in]{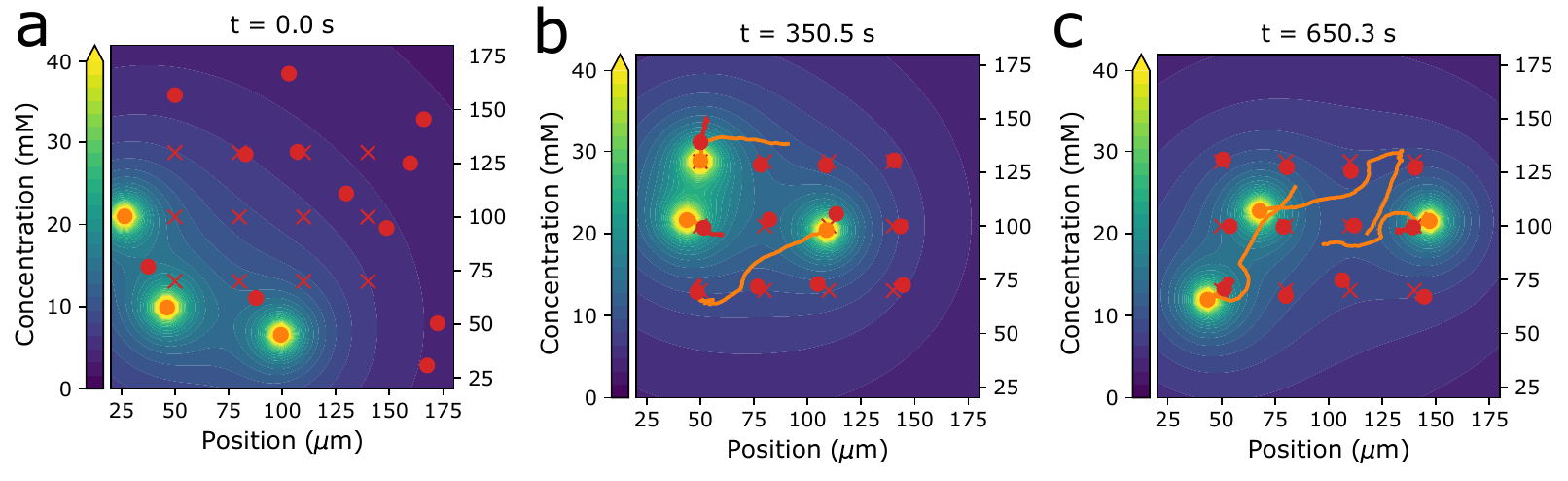}
  \caption{Three herders (orange) are used to move 12 followers (red) to their target positions (red x's).
Plot (a) shows the random initial condition. 
Plot (b) shows the particles after the herder has moved each particle to within $d_{\mathrm{tol}}$ of their targets. Plot (c) shows after it maintained them there for 5 minutes.
}
  \label{fig-ThreeHerders}
\end{figure*}

The BD equations and control laws derived in this paper assume a single herder, but they only need slight modifications to the concentration profile to model multiple herders. 
With multiple herders, we neglect the diffusiophoresis felt between herders. 
With this assumption, the BD equation for each herder is still given by Equation~\eqref{eq-brownianDynamics3} and the BD equation for each follower is still given by Equation~\eqref{eq-brownianDynamics2}.
However, $\nabla C_s$ is now evaluated as the sum of the solute gradient (Equation~\eqref{eq-concentrationgradient}) produced by each herder, or
\begin{equation}
\label{eq-concentrationgradientMultiple}
    \nabla C_s(\bm{X}) \approx  \sum_i \frac{-g_{h} (\bm{X}-\bm{y}_i) } {4\pi D_s ||\bm{X}-\bm{y}_i||^3} \textrm{.}
\end{equation}

The control algorithm developed in Section~\ref{sec-control} can be applied to each herder independently, with two modifications.
These modifications prevent the herders from getting too close to each other and causing the followers to group together, which was a common failure condition before these changes.
First, the switching strategy was modified by adding the constraint that a herder could not chase a follower located within $d_{\mathrm{approach}}$ of another herder.
Second, the GVF for each herder was modified to consider an area around each of the other herders as an obstacle to avoid, as detailed in the Supporting Information.

In Figure~\ref{fig-ThreeHerders}, three herders move 12 followers onto targets arranged on a regular lattice, using parameters from Table~\ref{table-table1}.
This example illustrates that scale-up to multiple herders is possible.
With the modifications mentioned above to ensure that herders never get too close together, we quickly and accurately arrange the particles on a lattice.
The full simulation is shown in Video~3, and it took 370 seconds to move all followers within $d_{\mathrm{tol}}$ of their targets. In 100 simulations with different initial conditions, the average time was 430 seconds, with a standard deviation of 130 seconds. 

We note that Equations~\eqref{eq-stabilityresultfull} and~\eqref{eq-browniancheck3}, which we used to predict the number of particles that can be steered by a single herder, do not apply to these multiple herder simulations.
The analysis that led to those bounds was based on a single herder.
However, while the numerical predictions are no longer valid, we expect the general principles that underlie Equations~\eqref{eq-stabilityresultfull} and~\eqref{eq-browniancheck3}---that herding is limited by Brownian motion and unwanted attraction---still apply to the multiple herder case.

Finally, it might be reasonable to attempt to make each herder steer a subset of particles so that no herder has to move too far, and so that scale-up is more intuitive.
However, initial attempts with this approach were unsuccessful because particles on the edges of the target pattern were attracted to the middle too strongly.
Particles near the edges need more attention than particles in the middle, so future herding strategies could focus on how to more efficiently partition herders and followers so that scale-up is intuitive and effective.

\section{Conclusion} \label{sec-conclusions}

In summary, we have used an externally-steered reactive particle to place passive particles on target positions, in a process we have dubbed chemical herding.
We did this using a control law divided in three parts: First, a switching strategy was employed and the optimal herder position was calculated.
Second, a GVF was used to determine the herder trajectory.
And third, an actuator-specific controller was needed to find the actuator values (e.g., electrode voltages) to make the herder move.
Using Lyapunov stability theory, we derived a bound on the number of particles that could be steered using a single herder, by comparing the desired attraction between the herder and a chased particle with the undesired attraction between the herder and an unchased particle. 
We added to this another bound due to Brownian motion and found the range of parameters in which particle steering is viable. 

Simulations were performed to validate the bounds we derived, and it was discovered that while these bounds are conservative, they capture the qualitative behavior of how parameters such as the radius of the followers cannot be either too large or too small for chemical herding to work. 
We conclude that chemical herding is viable for a narrow range of particle sizes, which, for the parameters chosen in our simulations, is roughly between 1 and 10$\mu$m. 
These sizes roughly correspond to the sizes of particles that are small enough to be considered colloids but large enough to be viewed through an optical microscope. 

Finally, multiple herders were used in tandem to demonstrate that chemical herding is a viable means to create a multiplicative factor on the number of particles that can be moved using top-down single-particle steering methods.
Chemical herding shows promise as a means to facilitate the precise, local control of a large number of particles using top-down methods and has great potential for the creation of dynamically configurable colloidal materials. 

\bibliography{refs} 

\begin{thebibliography}{0}%
\makeatletter
\providecommand \@ifxundefined [1]{%
 \@ifx{#1\undefined}
}%
\providecommand \@ifnum [1]{%
 \ifnum #1\expandafter \@firstoftwo
 \else \expandafter \@secondoftwo
 \fi
}%
\providecommand \@ifx [1]{%
 \ifx #1\expandafter \@firstoftwo
 \else \expandafter \@secondoftwo
 \fi
}%
\providecommand \natexlab [1]{#1}%
\providecommand \enquote  [1]{``#1''}%
\providecommand \bibnamefont  [1]{#1}%
\providecommand \bibfnamefont [1]{#1}%
\providecommand \citenamefont [1]{#1}%
\providecommand \href@noop [0]{\@secondoftwo}%
\providecommand \href [0]{\begingroup \@sanitize@url \@href}%
\providecommand \@href[1]{\@@startlink{#1}\@@href}%
\providecommand \@@href[1]{\endgroup#1\@@endlink}%
\providecommand \@sanitize@url [0]{\catcode `\\12\catcode `\$12\catcode
  `\&12\catcode `\#12\catcode `\^12\catcode `\_12\catcode `\%12\relax}%
\providecommand \@@startlink[1]{}%
\providecommand \@@endlink[0]{}%
\providecommand \url  [0]{\begingroup\@sanitize@url \@url }%
\providecommand \@url [1]{\endgroup\@href {#1}{\urlprefix }}%
\providecommand \urlprefix  [0]{URL }%
\providecommand \Eprint [0]{\href }%
\providecommand \doibase [0]{https://doi.org/}%
\providecommand \selectlanguage [0]{\@gobble}%
\providecommand \bibinfo  [0]{\@secondoftwo}%
\providecommand \bibfield  [0]{\@secondoftwo}%
\providecommand \translation [1]{[#1]}%
\providecommand \BibitemOpen [0]{}%
\providecommand \bibitemStop [0]{}%
\providecommand \bibitemNoStop [0]{.\EOS\space}%
\providecommand \EOS [0]{\spacefactor3000\relax}%
\providecommand \BibitemShut  [1]{\csname bibitem#1\endcsname}%
\let\auto@bib@innerbib\@empty
\end{thebibliography}%


\begin{thebibliography}{46}%
\makeatletter
\providecommand \@ifxundefined [1]{%
 \@ifx{#1\undefined}
}%
\providecommand \@ifnum [1]{%
 \ifnum #1\expandafter \@firstoftwo
 \else \expandafter \@secondoftwo
 \fi
}%
\providecommand \@ifx [1]{%
 \ifx #1\expandafter \@firstoftwo
 \else \expandafter \@secondoftwo
 \fi
}%
\providecommand \natexlab [1]{#1}%
\providecommand \enquote  [1]{``#1''}%
\providecommand \bibnamefont  [1]{#1}%
\providecommand \bibfnamefont [1]{#1}%
\providecommand \citenamefont [1]{#1}%
\providecommand \href@noop [0]{\@secondoftwo}%
\providecommand \href [0]{\begingroup \@sanitize@url \@href}%
\providecommand \@href[1]{\@@startlink{#1}\@@href}%
\providecommand \@@href[1]{\endgroup#1\@@endlink}%
\providecommand \@sanitize@url [0]{\catcode `\\12\catcode `\$12\catcode
  `\&12\catcode `\#12\catcode `\^12\catcode `\_12\catcode `\%12\relax}%
\providecommand \@@startlink[1]{}%
\providecommand \@@endlink[0]{}%
\providecommand \url  [0]{\begingroup\@sanitize@url \@url }%
\providecommand \@url [1]{\endgroup\@href {#1}{\urlprefix }}%
\providecommand \urlprefix  [0]{URL }%
\providecommand \Eprint [0]{\href }%
\providecommand \doibase [0]{https://doi.org/}%
\providecommand \selectlanguage [0]{\@gobble}%
\providecommand \bibinfo  [0]{\@secondoftwo}%
\providecommand \bibfield  [0]{\@secondoftwo}%
\providecommand \translation [1]{[#1]}%
\providecommand \BibitemOpen [0]{}%
\providecommand \bibitemStop [0]{}%
\providecommand \bibitemNoStop [0]{.\EOS\space}%
\providecommand \EOS [0]{\spacefactor3000\relax}%
\providecommand \BibitemShut  [1]{\csname bibitem#1\endcsname}%
\let\auto@bib@innerbib\@empty
\bibitem [{\citenamefont {Aubret}\ \emph {et~al.}(2018)\citenamefont {Aubret},
  \citenamefont {Youssef}, \citenamefont {Sacanna},\ and\ \citenamefont
  {Palacci}}]{Aubret2018}%
  \BibitemOpen
  \bibfield  {author} {\bibinfo {author} {\bibfnamefont {A.}~\bibnamefont
  {Aubret}}, \bibinfo {author} {\bibfnamefont {M.}~\bibnamefont {Youssef}},
  \bibinfo {author} {\bibfnamefont {S.}~\bibnamefont {Sacanna}},\ and\ \bibinfo
  {author} {\bibfnamefont {J.}~\bibnamefont {Palacci}},\ }\bibfield  {title}
  {\bibinfo {title} {{Targeted assembly and synchronization of self-spinning
  microgears}},\ }\href {https://doi.org/10.1038/s41567-018-0227-4} {\bibfield
  {journal} {\bibinfo  {journal} {Nature Physics}\ }\textbf {\bibinfo {volume}
  {14}},\ \bibinfo {pages} {1114} (\bibinfo {year} {2018})}\BibitemShut
  {NoStop}%
\bibitem [{\citenamefont {Aubret}\ \emph {et~al.}(2021)\citenamefont {Aubret},
  \citenamefont {Martinet},\ and\ \citenamefont {Palacci}}]{Aubret2021}%
  \BibitemOpen
  \bibfield  {author} {\bibinfo {author} {\bibfnamefont {A.}~\bibnamefont
  {Aubret}}, \bibinfo {author} {\bibfnamefont {Q.}~\bibnamefont {Martinet}},\
  and\ \bibinfo {author} {\bibfnamefont {J.}~\bibnamefont {Palacci}},\
  }\bibfield  {title} {\bibinfo {title} {{Metamachines of pluripotent
  colloids}},\ }\href {https://doi.org/10.1038/s41467-021-26699-6} {\bibfield
  {journal} {\bibinfo  {journal} {Nature Communications}\ }\textbf {\bibinfo
  {volume} {12}},\ \bibinfo {pages} {1} (\bibinfo {year} {2021})}\BibitemShut
  {NoStop}%
\bibitem [{\citenamefont {Soto}\ \emph {et~al.}(2022)\citenamefont {Soto},
  \citenamefont {Karshalev}, \citenamefont {Zhang}, \citenamefont {{Esteban
  Fernandez de Avila}}, \citenamefont {Nourhani},\ and\ \citenamefont
  {Wang}}]{Soto2022}%
  \BibitemOpen
  \bibfield  {author} {\bibinfo {author} {\bibfnamefont {F.}~\bibnamefont
  {Soto}}, \bibinfo {author} {\bibfnamefont {E.}~\bibnamefont {Karshalev}},
  \bibinfo {author} {\bibfnamefont {F.}~\bibnamefont {Zhang}}, \bibinfo
  {author} {\bibfnamefont {B.}~\bibnamefont {{Esteban Fernandez de Avila}}},
  \bibinfo {author} {\bibfnamefont {A.}~\bibnamefont {Nourhani}},\ and\
  \bibinfo {author} {\bibfnamefont {J.}~\bibnamefont {Wang}},\ }\bibfield
  {title} {\bibinfo {title} {{Smart Materials for Microrobots}},\ }\href
  {https://doi.org/10.1021/acs.chemrev.0c00999} {\bibfield  {journal} {\bibinfo
   {journal} {Chemical Reviews}\ }\textbf {\bibinfo {volume} {122}},\ \bibinfo
  {pages} {5365} (\bibinfo {year} {2022})}\BibitemShut {NoStop}%
\bibitem [{\citenamefont {Xie}\ \emph {et~al.}(2019)\citenamefont {Xie},
  \citenamefont {Sun}, \citenamefont {Fan}, \citenamefont {Lin}, \citenamefont
  {Chen}, \citenamefont {Wang}, \citenamefont {Dong},\ and\ \citenamefont
  {He}}]{Xie2019}%
  \BibitemOpen
  \bibfield  {author} {\bibinfo {author} {\bibfnamefont {H.}~\bibnamefont
  {Xie}}, \bibinfo {author} {\bibfnamefont {M.}~\bibnamefont {Sun}}, \bibinfo
  {author} {\bibfnamefont {X.}~\bibnamefont {Fan}}, \bibinfo {author}
  {\bibfnamefont {Z.}~\bibnamefont {Lin}}, \bibinfo {author} {\bibfnamefont
  {W.}~\bibnamefont {Chen}}, \bibinfo {author} {\bibfnamefont {L.}~\bibnamefont
  {Wang}}, \bibinfo {author} {\bibfnamefont {L.}~\bibnamefont {Dong}},\ and\
  \bibinfo {author} {\bibfnamefont {Q.}~\bibnamefont {He}},\ }\bibfield
  {title} {\bibinfo {title} {{Reconfigurable magnetic microrobot swarm:
  Multimode transformation, locomotion, and manipulation}},\ }\href
  {https://doi.org/10.1126/scirobotics.aav8006} {\bibfield  {journal} {\bibinfo
   {journal} {Science Robotics}\ }\textbf {\bibinfo {volume} {4}},\ \bibinfo
  {pages} {1} (\bibinfo {year} {2019})}\BibitemShut {NoStop}%
\bibitem [{\citenamefont {Palagi}\ \emph {et~al.}(2019)\citenamefont {Palagi},
  \citenamefont {Singh},\ and\ \citenamefont {Fischer}}]{Palagi2019}%
  \BibitemOpen
  \bibfield  {author} {\bibinfo {author} {\bibfnamefont {S.}~\bibnamefont
  {Palagi}}, \bibinfo {author} {\bibfnamefont {D.~P.}\ \bibnamefont {Singh}},\
  and\ \bibinfo {author} {\bibfnamefont {P.}~\bibnamefont {Fischer}},\
  }\bibfield  {title} {\bibinfo {title} {{Light-Controlled Micromotors and Soft
  Microrobots}},\ }\href {https://doi.org/10.1002/adom.201900370} {\bibfield
  {journal} {\bibinfo  {journal} {Advanced Optical Materials}\ }\textbf
  {\bibinfo {volume} {7}},\ \bibinfo {pages} {1} (\bibinfo {year}
  {2019})}\BibitemShut {NoStop}%
\bibitem [{\citenamefont {Liljestr{\"{o}}m}\ \emph {et~al.}(2019)\citenamefont
  {Liljestr{\"{o}}m}, \citenamefont {Chen}, \citenamefont {Dommersnes},
  \citenamefont {Fossum},\ and\ \citenamefont
  {Gr{\"{o}}schel}}]{Liljestrom2019}%
  \BibitemOpen
  \bibfield  {author} {\bibinfo {author} {\bibfnamefont {V.}~\bibnamefont
  {Liljestr{\"{o}}m}}, \bibinfo {author} {\bibfnamefont {C.}~\bibnamefont
  {Chen}}, \bibinfo {author} {\bibfnamefont {P.}~\bibnamefont {Dommersnes}},
  \bibinfo {author} {\bibfnamefont {J.~O.}\ \bibnamefont {Fossum}},\ and\
  \bibinfo {author} {\bibfnamefont {A.~H.}\ \bibnamefont {Gr{\"{o}}schel}},\
  }\bibfield  {title} {\bibinfo {title} {{Active structuring of colloids
  through field-driven self-assembly}},\ }\href
  {https://doi.org/10.1016/j.cocis.2018.10.008} {\bibfield  {journal} {\bibinfo
   {journal} {Current Opinion in Colloid and Interface Science}\ }\textbf
  {\bibinfo {volume} {40}},\ \bibinfo {pages} {25} (\bibinfo {year}
  {2019})}\BibitemShut {NoStop}%
\bibitem [{\citenamefont {Fu}\ \emph {et~al.}(2017)\citenamefont {Fu},
  \citenamefont {Chen}, \citenamefont {Zhao}, \citenamefont {Wang},
  \citenamefont {Shang}, \citenamefont {Gu},\ and\ \citenamefont
  {Zhao}}]{Fu2017}%
  \BibitemOpen
  \bibfield  {author} {\bibinfo {author} {\bibfnamefont {F.}~\bibnamefont
  {Fu}}, \bibinfo {author} {\bibfnamefont {Z.}~\bibnamefont {Chen}}, \bibinfo
  {author} {\bibfnamefont {Z.}~\bibnamefont {Zhao}}, \bibinfo {author}
  {\bibfnamefont {H.}~\bibnamefont {Wang}}, \bibinfo {author} {\bibfnamefont
  {L.}~\bibnamefont {Shang}}, \bibinfo {author} {\bibfnamefont
  {Z.}~\bibnamefont {Gu}},\ and\ \bibinfo {author} {\bibfnamefont
  {Y.}~\bibnamefont {Zhao}},\ }\bibfield  {title} {\bibinfo {title}
  {{Bio-inspired self-healing structural color hydrogel}},\ }\href
  {https://doi.org/10.1073/pnas.1703616114} {\bibfield  {journal} {\bibinfo
  {journal} {Proceedings of the National Academy of Sciences of the United
  States of America}\ }\textbf {\bibinfo {volume} {114}},\ \bibinfo {pages}
  {5900} (\bibinfo {year} {2017})}\BibitemShut {NoStop}%
\bibitem [{\citenamefont {Zhang}\ \emph {et~al.}(2020)\citenamefont {Zhang},
  \citenamefont {Bu}, \citenamefont {Yip}, \citenamefont {Liang},\ and\
  \citenamefont {Ho}}]{Zhang2020}%
  \BibitemOpen
  \bibfield  {author} {\bibinfo {author} {\bibfnamefont {H.}~\bibnamefont
  {Zhang}}, \bibinfo {author} {\bibfnamefont {X.}~\bibnamefont {Bu}}, \bibinfo
  {author} {\bibfnamefont {S.}~\bibnamefont {Yip}}, \bibinfo {author}
  {\bibfnamefont {X.}~\bibnamefont {Liang}},\ and\ \bibinfo {author}
  {\bibfnamefont {J.~C.}\ \bibnamefont {Ho}},\ }\bibfield  {title} {\bibinfo
  {title} {{Self‐Assembly of Colloidal Particles for Fabrication of
  Structural Color Materials toward Advanced Intelligent Systems}},\ }\href
  {https://doi.org/10.1002/aisy.201900085} {\bibfield  {journal} {\bibinfo
  {journal} {Advanced Intelligent Systems}\ }\textbf {\bibinfo {volume} {2}},\
  \bibinfo {pages} {1900085} (\bibinfo {year} {2020})}\BibitemShut {NoStop}%
\bibitem [{\citenamefont {Yadav}\ \emph {et~al.}(2020)\citenamefont {Yadav},
  \citenamefont {Sharma},\ and\ \citenamefont {Kumar}}]{Yadav2020}%
  \BibitemOpen
  \bibfield  {author} {\bibinfo {author} {\bibfnamefont {S.}~\bibnamefont
  {Yadav}}, \bibinfo {author} {\bibfnamefont {A.~K.}\ \bibnamefont {Sharma}},\
  and\ \bibinfo {author} {\bibfnamefont {P.}~\bibnamefont {Kumar}},\ }\bibfield
   {title} {\bibinfo {title} {{Nanoscale Self-Assembly for Therapeutic
  Delivery}},\ }\href {https://doi.org/10.3389/fbioe.2020.00127} {\bibfield
  {journal} {\bibinfo  {journal} {Front. Bioeng. Biotechnol.}\ }\textbf
  {\bibinfo {volume} {8}},\ \bibinfo {pages} {1} (\bibinfo {year}
  {2020})}\BibitemShut {NoStop}%
\bibitem [{\citenamefont {Gao}\ and\ \citenamefont {Wang}(2014)}]{Gao2014}%
  \BibitemOpen
  \bibfield  {author} {\bibinfo {author} {\bibfnamefont {W.}~\bibnamefont
  {Gao}}\ and\ \bibinfo {author} {\bibfnamefont {J.}~\bibnamefont {Wang}},\
  }\bibfield  {title} {\bibinfo {title} {Synthetic micro/nanomotors in drug
  delivery},\ }\href@noop {} {\bibfield  {journal} {\bibinfo  {journal}
  {Nanoscale}\ }\textbf {\bibinfo {volume} {6}},\ \bibinfo {pages} {10486}
  (\bibinfo {year} {2014})}\BibitemShut {NoStop}%
\bibitem [{\citenamefont {Singh}\ and\ \citenamefont
  {Moran}(2021)}]{Singh2021}%
  \BibitemOpen
  \bibfield  {author} {\bibinfo {author} {\bibfnamefont {S.}~\bibnamefont
  {Singh}}\ and\ \bibinfo {author} {\bibfnamefont {J.}~\bibnamefont {Moran}},\
  }\bibfield  {title} {\bibinfo {title} {{Autonomously Propelled Colloids for
  Penetration and Payload Delivery in Complex Extracellular Matrices}},\ }\href
  {https://doi.org/10.3390/mi12101216} {\bibfield  {journal} {\bibinfo
  {journal} {Micromachines}\ }\textbf {\bibinfo {volume} {12}},\ \bibinfo
  {pages} {1216} (\bibinfo {year} {2021})}\BibitemShut {NoStop}%
\bibitem [{\citenamefont {Isaacoff}\ and\ \citenamefont
  {Brown}(2017)}]{Isaacoff2017}%
  \BibitemOpen
  \bibfield  {author} {\bibinfo {author} {\bibfnamefont {B.~P.}\ \bibnamefont
  {Isaacoff}}\ and\ \bibinfo {author} {\bibfnamefont {K.~A.}\ \bibnamefont
  {Brown}},\ }\bibfield  {title} {\bibinfo {title} {{Progress in Top-Down
  Control of Bottom-Up Assembly}},\ }\href
  {https://doi.org/10.1021/acs.nanolett.7b04479} {\bibfield  {journal}
  {\bibinfo  {journal} {Nano Letters}\ }\textbf {\bibinfo {volume} {17}},\
  \bibinfo {pages} {6508} (\bibinfo {year} {2017})}\BibitemShut {NoStop}%
\bibitem [{\citenamefont {Armani}\ \emph {et~al.}(2006)\citenamefont {Armani},
  \citenamefont {Chaudhary}, \citenamefont {Probst},\ and\ \citenamefont
  {Shapiro}}]{Armani2006}%
  \BibitemOpen
  \bibfield  {author} {\bibinfo {author} {\bibfnamefont {M.~D.}\ \bibnamefont
  {Armani}}, \bibinfo {author} {\bibfnamefont {S.~V.}\ \bibnamefont
  {Chaudhary}}, \bibinfo {author} {\bibfnamefont {R.}~\bibnamefont {Probst}},\
  and\ \bibinfo {author} {\bibfnamefont {B.}~\bibnamefont {Shapiro}},\
  }\bibfield  {title} {\bibinfo {title} {{Using feedback control of microflows
  to independently steer multiple particles}},\ }\href
  {https://doi.org/10.1109/JMEMS.2006.878863} {\bibfield  {journal} {\bibinfo
  {journal} {J. Microelectromech. Syst.}\ }\textbf {\bibinfo {volume} {15}},\
  \bibinfo {pages} {945} (\bibinfo {year} {2006})}\BibitemShut {NoStop}%
\bibitem [{\citenamefont {Cummins}\ \emph {et~al.}(2013)\citenamefont
  {Cummins}, \citenamefont {Probst},\ and\ \citenamefont
  {Shapiro}}]{Cummins2013}%
  \BibitemOpen
  \bibfield  {author} {\bibinfo {author} {\bibfnamefont {Z.}~\bibnamefont
  {Cummins}}, \bibinfo {author} {\bibfnamefont {R.}~\bibnamefont {Probst}},\
  and\ \bibinfo {author} {\bibfnamefont {B.}~\bibnamefont {Shapiro}},\
  }\bibfield  {title} {\bibinfo {title} {{Electrokinetic tweezing:
  three-dimensional manipulation of microparticles by real-time imaging and
  flow control}},\ }\href {https://doi.org/10.1039/c3lc50674f} {\bibfield
  {journal} {\bibinfo  {journal} {Lab on a Chip}\ }\textbf {\bibinfo {volume}
  {13}},\ \bibinfo {pages} {4040} (\bibinfo {year} {2013})}\BibitemShut
  {NoStop}%
\bibitem [{\citenamefont {Matei}\ \emph {et~al.}(2019)\citenamefont {Matei},
  \citenamefont {Nelaturi}, \citenamefont {Chow}, \citenamefont {Lu},
  \citenamefont {Bert},\ and\ \citenamefont {Crawford}}]{Matei2019}%
  \BibitemOpen
  \bibfield  {author} {\bibinfo {author} {\bibfnamefont {I.}~\bibnamefont
  {Matei}}, \bibinfo {author} {\bibfnamefont {S.}~\bibnamefont {Nelaturi}},
  \bibinfo {author} {\bibfnamefont {E.~M.}\ \bibnamefont {Chow}}, \bibinfo
  {author} {\bibfnamefont {J.~P.}\ \bibnamefont {Lu}}, \bibinfo {author}
  {\bibfnamefont {J.~A.}\ \bibnamefont {Bert}},\ and\ \bibinfo {author}
  {\bibfnamefont {L.~S.}\ \bibnamefont {Crawford}},\ }\bibfield  {title}
  {\bibinfo {title} {{Micro-Scale Chiplets Position Control}},\ }\href
  {https://doi.org/10.1109/JMEMS.2019.2914045} {\bibfield  {journal} {\bibinfo
  {journal} {Journal of Microelectromechanical Systems}\ }\textbf {\bibinfo
  {volume} {28}},\ \bibinfo {pages} {643} (\bibinfo {year} {2019})}\BibitemShut
  {NoStop}%
\bibitem [{\citenamefont {Shenoy}\ \emph {et~al.}(2016)\citenamefont {Shenoy},
  \citenamefont {Rao},\ and\ \citenamefont {Schroeder}}]{Shenoy2016}%
  \BibitemOpen
  \bibfield  {author} {\bibinfo {author} {\bibfnamefont {A.}~\bibnamefont
  {Shenoy}}, \bibinfo {author} {\bibfnamefont {C.~V.}\ \bibnamefont {Rao}},\
  and\ \bibinfo {author} {\bibfnamefont {C.~M.}\ \bibnamefont {Schroeder}},\
  }\bibfield  {title} {\bibinfo {title} {{Stokes trap for multiplexed particle
  manipulation and assembly using fluidics}},\ }\href
  {https://doi.org/10.1073/pnas.1525162113} {\bibfield  {journal} {\bibinfo
  {journal} {Proceedings of the National Academy of Sciences of the United
  States of America}\ }\textbf {\bibinfo {volume} {113}},\ \bibinfo {pages}
  {3976} (\bibinfo {year} {2016})}\BibitemShut {NoStop}%
\bibitem [{\citenamefont {Shenoy}\ \emph {et~al.}(2019)\citenamefont {Shenoy},
  \citenamefont {Kumar}, \citenamefont {Hilgenfeldt},\ and\ \citenamefont
  {Schroeder}}]{Shenoy2019}%
  \BibitemOpen
  \bibfield  {author} {\bibinfo {author} {\bibfnamefont {A.}~\bibnamefont
  {Shenoy}}, \bibinfo {author} {\bibfnamefont {D.}~\bibnamefont {Kumar}},
  \bibinfo {author} {\bibfnamefont {S.}~\bibnamefont {Hilgenfeldt}},\ and\
  \bibinfo {author} {\bibfnamefont {C.~M.}\ \bibnamefont {Schroeder}},\
  }\bibfield  {title} {\bibinfo {title} {{Flow Topology during Multiplexed
  Particle Manipulation Using a Stokes Trap}},\ }\href
  {https://doi.org/10.1103/PhysRevApplied.12.054010} {\bibfield  {journal}
  {\bibinfo  {journal} {Physical Review Applied}\ }\textbf {\bibinfo {volume}
  {12}},\ \bibinfo {pages} {1} (\bibinfo {year} {2019})}\BibitemShut {NoStop}%
\bibitem [{\citenamefont {Kumar}\ \emph {et~al.}(2019)\citenamefont {Kumar},
  \citenamefont {Shenoy}, \citenamefont {Li},\ and\ \citenamefont
  {Schroeder}}]{Kumar2019}%
  \BibitemOpen
  \bibfield  {author} {\bibinfo {author} {\bibfnamefont {D.}~\bibnamefont
  {Kumar}}, \bibinfo {author} {\bibfnamefont {A.}~\bibnamefont {Shenoy}},
  \bibinfo {author} {\bibfnamefont {S.}~\bibnamefont {Li}},\ and\ \bibinfo
  {author} {\bibfnamefont {C.~M.}\ \bibnamefont {Schroeder}},\ }\bibfield
  {title} {\bibinfo {title} {{Orientation control and nonlinear trajectory
  tracking of colloidal particles using microfluidics}},\ }\href
  {https://doi.org/10.1103/PhysRevFluids.4.114203} {\bibfield  {journal}
  {\bibinfo  {journal} {Physical Review Fluids}\ }\textbf {\bibinfo {volume}
  {4}},\ \bibinfo {pages} {1} (\bibinfo {year} {2019})},\ \Eprint
  {https://arxiv.org/abs/1907.08567} {1907.08567} \BibitemShut {NoStop}%
\bibitem [{\citenamefont {Gosse}\ and\ \citenamefont
  {Croquette}(2002)}]{Gosse2002}%
  \BibitemOpen
  \bibfield  {author} {\bibinfo {author} {\bibfnamefont {C.}~\bibnamefont
  {Gosse}}\ and\ \bibinfo {author} {\bibfnamefont {V.}~\bibnamefont
  {Croquette}},\ }\bibfield  {title} {\bibinfo {title} {{Magnetic Tweezers:
  Micromanipulation and Force Measurement at the Molecular Level}},\ }\href
  {https://doi.org/10.1016/S0006-3495(02)75672-5} {\bibfield  {journal}
  {\bibinfo  {journal} {Biophysical Journal}\ }\textbf {\bibinfo {volume}
  {82}},\ \bibinfo {pages} {3314} (\bibinfo {year} {2002})}\BibitemShut
  {NoStop}%
\bibitem [{\citenamefont {Ashkin}\ \emph {et~al.}(1986)\citenamefont {Ashkin},
  \citenamefont {Dziedzic}, \citenamefont {Bjorkholm},\ and\ \citenamefont
  {Chu}}]{Ashkin1986}%
  \BibitemOpen
  \bibfield  {author} {\bibinfo {author} {\bibfnamefont {A.}~\bibnamefont
  {Ashkin}}, \bibinfo {author} {\bibfnamefont {J.~M.}\ \bibnamefont
  {Dziedzic}}, \bibinfo {author} {\bibfnamefont {J.~E.}\ \bibnamefont
  {Bjorkholm}},\ and\ \bibinfo {author} {\bibfnamefont {S.}~\bibnamefont
  {Chu}},\ }\bibfield  {title} {\bibinfo {title} {{Observation of a single-beam
  gradient force optical trap for dielectric particles}},\ }\href
  {https://doi.org/10.1364/OL.11.000288} {\bibfield  {journal} {\bibinfo
  {journal} {Optics Letters}\ }\textbf {\bibinfo {volume} {11}},\ \bibinfo
  {pages} {288} (\bibinfo {year} {1986})}\BibitemShut {NoStop}%
\bibitem [{\citenamefont {Moffitt}\ \emph {et~al.}(2008)\citenamefont
  {Moffitt}, \citenamefont {Chemla}, \citenamefont {Smith},\ and\ \citenamefont
  {Bustamante}}]{Moffitt2008}%
  \BibitemOpen
  \bibfield  {author} {\bibinfo {author} {\bibfnamefont {J.~R.}\ \bibnamefont
  {Moffitt}}, \bibinfo {author} {\bibfnamefont {Y.~R.}\ \bibnamefont {Chemla}},
  \bibinfo {author} {\bibfnamefont {S.~B.}\ \bibnamefont {Smith}},\ and\
  \bibinfo {author} {\bibfnamefont {C.}~\bibnamefont {Bustamante}},\ }\bibfield
   {title} {\bibinfo {title} {{Recent Advances in Optical Tweezers}},\ }\href
  {https://doi.org/10.1146/annurev.biochem.77.043007.090225} {\bibfield
  {journal} {\bibinfo  {journal} {Annual Review of Biochemistry}\ }\textbf
  {\bibinfo {volume} {77}},\ \bibinfo {pages} {205} (\bibinfo {year}
  {2008})}\BibitemShut {NoStop}%
\bibitem [{\citenamefont {Marzo}\ and\ \citenamefont
  {Drinkwater}(2019)}]{Marzo2019}%
  \BibitemOpen
  \bibfield  {author} {\bibinfo {author} {\bibfnamefont {A.}~\bibnamefont
  {Marzo}}\ and\ \bibinfo {author} {\bibfnamefont {B.~W.}\ \bibnamefont
  {Drinkwater}},\ }\bibfield  {title} {\bibinfo {title} {{Holographic acoustic
  tweezers}},\ }\href {https://doi.org/10.1073/pnas.1813047115} {\bibfield
  {journal} {\bibinfo  {journal} {Proceedings of the National Academy of
  Sciences of the United States of America}\ }\textbf {\bibinfo {volume}
  {116}},\ \bibinfo {pages} {84} (\bibinfo {year} {2019})}\BibitemShut
  {NoStop}%
\bibitem [{\citenamefont {Nilsson}\ \emph {et~al.}(2009)\citenamefont
  {Nilsson}, \citenamefont {Evander}, \citenamefont {Hammarstr{\"{o}}m},\ and\
  \citenamefont {Laurell}}]{Nilsson2009}%
  \BibitemOpen
  \bibfield  {author} {\bibinfo {author} {\bibfnamefont {J.}~\bibnamefont
  {Nilsson}}, \bibinfo {author} {\bibfnamefont {M.}~\bibnamefont {Evander}},
  \bibinfo {author} {\bibfnamefont {B.}~\bibnamefont {Hammarstr{\"{o}}m}},\
  and\ \bibinfo {author} {\bibfnamefont {T.}~\bibnamefont {Laurell}},\
  }\bibfield  {title} {\bibinfo {title} {{Review of cell and particle trapping
  in microfluidic systems}},\ }\href
  {https://doi.org/10.1016/j.aca.2009.07.017} {\bibfield  {journal} {\bibinfo
  {journal} {Analytica Chimica Acta}\ }\textbf {\bibinfo {volume} {649}},\
  \bibinfo {pages} {141} (\bibinfo {year} {2009})}\BibitemShut {NoStop}%
\bibitem [{\citenamefont {Ding}\ \emph {et~al.}(2012)\citenamefont {Ding},
  \citenamefont {Lin}, \citenamefont {Kiraly}, \citenamefont {Yue},
  \citenamefont {Li}, \citenamefont {Chiang}, \citenamefont {Shi},
  \citenamefont {Benkovic},\ and\ \citenamefont {Huang}}]{Ding2012}%
  \BibitemOpen
  \bibfield  {author} {\bibinfo {author} {\bibfnamefont {X.}~\bibnamefont
  {Ding}}, \bibinfo {author} {\bibfnamefont {S.~C.~S.}\ \bibnamefont {Lin}},
  \bibinfo {author} {\bibfnamefont {B.}~\bibnamefont {Kiraly}}, \bibinfo
  {author} {\bibfnamefont {H.}~\bibnamefont {Yue}}, \bibinfo {author}
  {\bibfnamefont {S.}~\bibnamefont {Li}}, \bibinfo {author} {\bibfnamefont
  {I.~K.}\ \bibnamefont {Chiang}}, \bibinfo {author} {\bibfnamefont
  {J.}~\bibnamefont {Shi}}, \bibinfo {author} {\bibfnamefont {S.~J.}\
  \bibnamefont {Benkovic}},\ and\ \bibinfo {author} {\bibfnamefont {T.~J.}\
  \bibnamefont {Huang}},\ }\bibfield  {title} {\bibinfo {title} {{On-chip
  manipulation of single microparticles, cells, and organisms using surface
  acoustic waves}},\ }\href {https://doi.org/10.1073/pnas.1209288109}
  {\bibfield  {journal} {\bibinfo  {journal} {Proceedings of the National
  Academy of Sciences of the United States of America}\ }\textbf {\bibinfo
  {volume} {109}},\ \bibinfo {pages} {11105} (\bibinfo {year}
  {2012})}\BibitemShut {NoStop}%
\bibitem [{\citenamefont {Yang}\ and\ \citenamefont {Bevan}(2018)}]{Yang2018}%
  \BibitemOpen
  \bibfield  {author} {\bibinfo {author} {\bibfnamefont {Y.}~\bibnamefont
  {Yang}}\ and\ \bibinfo {author} {\bibfnamefont {M.~A.}\ \bibnamefont
  {Bevan}},\ }\bibfield  {title} {\bibinfo {title} {{Optimal Navigation of
  Self-Propelled Colloids}},\ }\href {https://doi.org/10.1021/acsnano.8b05371}
  {\bibfield  {journal} {\bibinfo  {journal} {ACS Nano}\ }\textbf {\bibinfo
  {volume} {12}},\ \bibinfo {pages} {10712} (\bibinfo {year}
  {2018})}\BibitemShut {NoStop}%
\bibitem [{\citenamefont {McDonald}\ \emph {et~al.}(2023)\citenamefont
  {McDonald}, \citenamefont {Peterson},\ and\ \citenamefont
  {Tree}}]{McDonald2023a}%
  \BibitemOpen
  \bibfield  {author} {\bibinfo {author} {\bibfnamefont {M.~N.}\ \bibnamefont
  {McDonald}}, \bibinfo {author} {\bibfnamefont {C.~K.}\ \bibnamefont
  {Peterson}},\ and\ \bibinfo {author} {\bibfnamefont {D.~R.}\ \bibnamefont
  {Tree}},\ }\bibfield  {title} {\bibinfo {title} {{Steering particles via
  micro-actuation of chemical gradients using model predictive control}},\
  }\href {https://doi.org/10.1063/5.0126690} {\bibfield  {journal} {\bibinfo
  {journal} {Biomicrofluidics}\ }\textbf {\bibinfo {volume} {17}},\ \bibinfo
  {pages} {014107} (\bibinfo {year} {2023})}\BibitemShut {NoStop}%
\bibitem [{\citenamefont {Kunche}\ \emph {et~al.}(2016)\citenamefont {Kunche},
  \citenamefont {Yan}, \citenamefont {Calof}, \citenamefont {Lowengrub},\ and\
  \citenamefont {Lander}}]{Kunche2016}%
  \BibitemOpen
  \bibfield  {author} {\bibinfo {author} {\bibfnamefont {S.}~\bibnamefont
  {Kunche}}, \bibinfo {author} {\bibfnamefont {H.}~\bibnamefont {Yan}},
  \bibinfo {author} {\bibfnamefont {A.~L.}\ \bibnamefont {Calof}}, \bibinfo
  {author} {\bibfnamefont {J.~S.}\ \bibnamefont {Lowengrub}},\ and\ \bibinfo
  {author} {\bibfnamefont {A.~D.}\ \bibnamefont {Lander}},\ }\bibfield  {title}
  {\bibinfo {title} {{Feedback, Lineages and Self-Organizing Morphogenesis}},\
  }\href {https://doi.org/10.1371/journal.pcbi.1004814} {\bibfield  {journal}
  {\bibinfo  {journal} {PLoS Comput. Biol.}\ }\textbf {\bibinfo {volume}
  {12}},\ \bibinfo {pages} {e1004814} (\bibinfo {year} {2016})}\BibitemShut
  {NoStop}%
\bibitem [{\citenamefont {Licitra}\ \emph {et~al.}(2017)\citenamefont
  {Licitra}, \citenamefont {Hutcheson}, \citenamefont {Doucette},\ and\
  \citenamefont {Dixon}}]{Licitra2017}%
  \BibitemOpen
  \bibfield  {author} {\bibinfo {author} {\bibfnamefont {R.~A.}\ \bibnamefont
  {Licitra}}, \bibinfo {author} {\bibfnamefont {Z.~D.}\ \bibnamefont
  {Hutcheson}}, \bibinfo {author} {\bibfnamefont {E.~A.}\ \bibnamefont
  {Doucette}},\ and\ \bibinfo {author} {\bibfnamefont {W.~E.}\ \bibnamefont
  {Dixon}},\ }\bibfield  {title} {\bibinfo {title} {{Single Agent Herding of
  n-Agents: A Switched Systems Approach}},\ }\href
  {https://doi.org/10.1016/j.ifacol.2017.08.2020} {\bibfield  {journal}
  {\bibinfo  {journal} {IFAC-PapersOnLine}\ }\textbf {\bibinfo {volume} {50}},\
  \bibinfo {pages} {14374} (\bibinfo {year} {2017})}\BibitemShut {NoStop}%
\bibitem [{\citenamefont {Licitra}\ \emph {et~al.}(2018)\citenamefont
  {Licitra}, \citenamefont {Bell}, \citenamefont {Doucette},\ and\
  \citenamefont {Dixon}}]{Licitra2018}%
  \BibitemOpen
  \bibfield  {author} {\bibinfo {author} {\bibfnamefont {R.~A.}\ \bibnamefont
  {Licitra}}, \bibinfo {author} {\bibfnamefont {Z.~I.}\ \bibnamefont {Bell}},
  \bibinfo {author} {\bibfnamefont {E.~A.}\ \bibnamefont {Doucette}},\ and\
  \bibinfo {author} {\bibfnamefont {W.~E.}\ \bibnamefont {Dixon}},\ }\bibfield
  {title} {\bibinfo {title} {{Single agent indirect herding of multiple
  targets: A switched adaptive control approach}},\ }\href
  {https://doi.org/10.1109/LCSYS.2017.2763968} {\bibfield  {journal} {\bibinfo
  {journal} {IEEE Control Systems Letters}\ }\textbf {\bibinfo {volume} {2}},\
  \bibinfo {pages} {127} (\bibinfo {year} {2018})}\BibitemShut {NoStop}%
\bibitem [{\citenamefont {Licitra}\ \emph {et~al.}(2019)\citenamefont
  {Licitra}, \citenamefont {Bell},\ and\ \citenamefont {DIxon}}]{Licitra2019}%
  \BibitemOpen
  \bibfield  {author} {\bibinfo {author} {\bibfnamefont {R.~A.}\ \bibnamefont
  {Licitra}}, \bibinfo {author} {\bibfnamefont {Z.~I.}\ \bibnamefont {Bell}},\
  and\ \bibinfo {author} {\bibfnamefont {W.~E.}\ \bibnamefont {DIxon}},\
  }\bibfield  {title} {\bibinfo {title} {{Single-Agent Indirect Herding of
  Multiple Targets with Uncertain Dynamics}},\ }\href
  {https://doi.org/10.1109/TRO.2019.2911799} {\bibfield  {journal} {\bibinfo
  {journal} {IEEE Transactions on Robotics}\ }\textbf {\bibinfo {volume}
  {35}},\ \bibinfo {pages} {847} (\bibinfo {year} {2019})}\BibitemShut
  {NoStop}%
\bibitem [{\citenamefont {Zhang}\ \emph {et~al.}(2021)\citenamefont {Zhang},
  \citenamefont {Mou}, \citenamefont {Wu}, \citenamefont {Song}, \citenamefont
  {Kauffman}, \citenamefont {Sen},\ and\ \citenamefont {Guan}}]{Zhang2021}%
  \BibitemOpen
  \bibfield  {author} {\bibinfo {author} {\bibfnamefont {J.}~\bibnamefont
  {Zhang}}, \bibinfo {author} {\bibfnamefont {F.}~\bibnamefont {Mou}}, \bibinfo
  {author} {\bibfnamefont {Z.}~\bibnamefont {Wu}}, \bibinfo {author}
  {\bibfnamefont {J.}~\bibnamefont {Song}}, \bibinfo {author} {\bibfnamefont
  {J.~E.}\ \bibnamefont {Kauffman}}, \bibinfo {author} {\bibfnamefont
  {A.}~\bibnamefont {Sen}},\ and\ \bibinfo {author} {\bibfnamefont
  {J.}~\bibnamefont {Guan}},\ }\bibfield  {title} {\bibinfo {title}
  {{Cooperative transport by flocking phototactic micromotors}},\ }\href
  {https://doi.org/10.1039/d1na00641j} {\bibfield  {journal} {\bibinfo
  {journal} {Nanoscale Advances}\ }\textbf {\bibinfo {volume} {3}},\ \bibinfo
  {pages} {6157} (\bibinfo {year} {2021})}\BibitemShut {NoStop}%
\bibitem [{\citenamefont {Singh}\ \emph {et~al.}(2017)\citenamefont {Singh},
  \citenamefont {Choudhury}, \citenamefont {Fischer},\ and\ \citenamefont
  {Mark}}]{Singh2017}%
  \BibitemOpen
  \bibfield  {author} {\bibinfo {author} {\bibfnamefont {D.~P.}\ \bibnamefont
  {Singh}}, \bibinfo {author} {\bibfnamefont {U.}~\bibnamefont {Choudhury}},
  \bibinfo {author} {\bibfnamefont {P.}~\bibnamefont {Fischer}},\ and\ \bibinfo
  {author} {\bibfnamefont {A.~G.}\ \bibnamefont {Mark}},\ }\bibfield  {title}
  {\bibinfo {title} {{Non-Equilibrium Assembly of Light-Activated Colloidal
  Mixtures}},\ }\href {https://doi.org/10.1002/adma.201701328} {\bibfield
  {journal} {\bibinfo  {journal} {Advanced Materials}\ }\textbf {\bibinfo
  {volume} {29}},\ \bibinfo {pages} {1} (\bibinfo {year} {2017})}\BibitemShut
  {NoStop}%
\bibitem [{\citenamefont {Dorfman}\ \emph {et~al.}(2014)\citenamefont
  {Dorfman}, \citenamefont {Gupta}, \citenamefont {Jain}, \citenamefont
  {Muralidhar},\ and\ \citenamefont {Tree}}]{Dorfman2014}%
  \BibitemOpen
  \bibfield  {author} {\bibinfo {author} {\bibfnamefont {K.~D.}\ \bibnamefont
  {Dorfman}}, \bibinfo {author} {\bibfnamefont {D.}~\bibnamefont {Gupta}},
  \bibinfo {author} {\bibfnamefont {A.}~\bibnamefont {Jain}}, \bibinfo {author}
  {\bibfnamefont {A.}~\bibnamefont {Muralidhar}},\ and\ \bibinfo {author}
  {\bibfnamefont {D.~R.}\ \bibnamefont {Tree}},\ }\bibfield  {title} {\bibinfo
  {title} {{Hydrodynamics of DNA confined in nanoslits and nanochannels}},\
  }\href {https://doi.org/10.1140/epjst/e2014-02326-4} {\bibfield  {journal}
  {\bibinfo  {journal} {Eur. Phys. J. Spec. Top.}\ }\textbf {\bibinfo {volume}
  {223}},\ \bibinfo {pages} {3179} (\bibinfo {year} {2014})}\BibitemShut
  {NoStop}%
\bibitem [{\citenamefont {{\"{O}}ttinger}(1996)}]{Ottinger1996}%
  \BibitemOpen
  \bibfield  {author} {\bibinfo {author} {\bibfnamefont {H.~C.}\ \bibnamefont
  {{\"{O}}ttinger}},\ }\href {https://doi.org/10.1007/978-3-642-58290-5} {\emph
  {\bibinfo {title} {{Stochastic Processes in Polymeric Fluids}}}}\ (\bibinfo
  {publisher} {Springer Berlin Heidelberg},\ \bibinfo {year}
  {1996})\BibitemShut {NoStop}%
\bibitem [{\citenamefont {Anderson}(1989)}]{Anderson1989}%
  \BibitemOpen
  \bibfield  {author} {\bibinfo {author} {\bibfnamefont {J.}~\bibnamefont
  {Anderson}},\ }\bibfield  {title} {\bibinfo {title} {{Colloid Transport By
  Interfacial Forces}},\ }\href {https://doi.org/10.1146/annurev.fluid.21.1.61}
  {\bibfield  {journal} {\bibinfo  {journal} {Annual Review of Fluid
  Mechanics}\ }\textbf {\bibinfo {volume} {21}},\ \bibinfo {pages} {61}
  (\bibinfo {year} {1989})}\BibitemShut {NoStop}%
\bibitem [{\citenamefont {Heyes}\ and\ \citenamefont
  {Melrose}(1993)}]{Heyes1993}%
  \BibitemOpen
  \bibfield  {author} {\bibinfo {author} {\bibfnamefont {D.~M.}\ \bibnamefont
  {Heyes}}\ and\ \bibinfo {author} {\bibfnamefont {J.~R.}\ \bibnamefont
  {Melrose}},\ }\bibfield  {title} {\bibinfo {title} {{Brownian dynamics
  simulations of model hard-sphere suspensions}},\ }\href
  {https://doi.org/10.1016/0377-0257(93)80001-R} {\bibfield  {journal}
  {\bibinfo  {journal} {Journal of Non-Newtonian Fluid Mechanics}\ }\textbf
  {\bibinfo {volume} {46}},\ \bibinfo {pages} {1} (\bibinfo {year}
  {1993})}\BibitemShut {NoStop}%
\bibitem [{\citenamefont {Haberman}(2004)}]{Haberman2004}%
  \BibitemOpen
  \bibfield  {author} {\bibinfo {author} {\bibfnamefont {R.}~\bibnamefont
  {Haberman}},\ }\href@noop {} {\emph {\bibinfo {title} {{Applied Partial
  Differential Equations: with Fourier Series and Boundary Value Problems (4th
  ed.)}}}}\ (\bibinfo  {publisher} {Prentice Hall},\ \bibinfo {year}
  {2004})\BibitemShut {NoStop}%
\bibitem [{\citenamefont {Liebchen}\ \emph {et~al.}(2017)\citenamefont
  {Liebchen}, \citenamefont {Marenduzzo},\ and\ \citenamefont
  {Cates}}]{Liebchen2017}%
  \BibitemOpen
  \bibfield  {author} {\bibinfo {author} {\bibfnamefont {B.}~\bibnamefont
  {Liebchen}}, \bibinfo {author} {\bibfnamefont {D.}~\bibnamefont
  {Marenduzzo}},\ and\ \bibinfo {author} {\bibfnamefont {M.~E.}\ \bibnamefont
  {Cates}},\ }\bibfield  {title} {\bibinfo {title} {{Phoretic Interactions
  Generically Induce Dynamic Clusters and Wave Patterns in Active Colloids}},\
  }\href {https://doi.org/10.1103/PhysRevLett.118.268001} {\bibfield  {journal}
  {\bibinfo  {journal} {Physical Review Letters}\ }\textbf {\bibinfo {volume}
  {118}},\ \bibinfo {pages} {1} (\bibinfo {year} {2017})}\BibitemShut {NoStop}%
\bibitem [{\citenamefont {Liebchen}\ and\ \citenamefont
  {L{\"{o}}wen}(2019)}]{Liebchen2019}%
  \BibitemOpen
  \bibfield  {author} {\bibinfo {author} {\bibfnamefont {B.}~\bibnamefont
  {Liebchen}}\ and\ \bibinfo {author} {\bibfnamefont {H.}~\bibnamefont
  {L{\"{o}}wen}},\ }\bibfield  {title} {\bibinfo {title} {{Which interactions
  dominate in active colloids?}},\ }\bibfield  {journal} {\bibinfo  {journal}
  {Journal of Chemical Physics}\ }\textbf {\bibinfo {volume} {150}},\ \href
  {https://doi.org/10.1063/1.5082284} {10.1063/1.5082284} (\bibinfo {year}
  {2019})\BibitemShut {NoStop}%
\bibitem [{\citenamefont {Zhou}\ \emph {et~al.}(2021)\citenamefont {Zhou},
  \citenamefont {Wang}, \citenamefont {Xian}, \citenamefont {Shah},
  \citenamefont {Li}, \citenamefont {Lin},\ and\ \citenamefont
  {Gao}}]{Zhou2021}%
  \BibitemOpen
  \bibfield  {author} {\bibinfo {author} {\bibfnamefont {X.}~\bibnamefont
  {Zhou}}, \bibinfo {author} {\bibfnamefont {S.}~\bibnamefont {Wang}}, \bibinfo
  {author} {\bibfnamefont {L.}~\bibnamefont {Xian}}, \bibinfo {author}
  {\bibfnamefont {Z.~H.}\ \bibnamefont {Shah}}, \bibinfo {author}
  {\bibfnamefont {Y.}~\bibnamefont {Li}}, \bibinfo {author} {\bibfnamefont
  {G.}~\bibnamefont {Lin}},\ and\ \bibinfo {author} {\bibfnamefont
  {Y.}~\bibnamefont {Gao}},\ }\bibfield  {title} {\bibinfo {title} {{Ionic
  Effects in Ionic Diffusiophoresis in Chemically Driven Active Colloids}},\
  }\href {https://doi.org/10.1103/PhysRevLett.127.168001} {\bibfield  {journal}
  {\bibinfo  {journal} {Physical Review Letters}\ }\textbf {\bibinfo {volume}
  {127}},\ \bibinfo {pages} {168001} (\bibinfo {year} {2021})}\BibitemShut
  {NoStop}%
\bibitem [{\citenamefont {Wilhelm}\ and\ \citenamefont
  {Clem}(2019)}]{Wilhelm2019}%
  \BibitemOpen
  \bibfield  {author} {\bibinfo {author} {\bibfnamefont {J.~P.}\ \bibnamefont
  {Wilhelm}}\ and\ \bibinfo {author} {\bibfnamefont {G.}~\bibnamefont {Clem}},\
  }\bibfield  {title} {\bibinfo {title} {{Vector field UAV guidance for path
  following and obstacle avoidance with minimal deviation}},\ }\href
  {https://doi.org/10.2514/1.G004053} {\bibfield  {journal} {\bibinfo
  {journal} {Journal of Guidance, Control, and Dynamics}\ }\textbf {\bibinfo
  {volume} {42}},\ \bibinfo {pages} {1848} (\bibinfo {year}
  {2019})}\BibitemShut {NoStop}%
\bibitem [{\citenamefont {Chaudhary}\ and\ \citenamefont
  {Shapiro}(2006)}]{Chaudhary2006}%
  \BibitemOpen
  \bibfield  {author} {\bibinfo {author} {\bibfnamefont {S.}~\bibnamefont
  {Chaudhary}}\ and\ \bibinfo {author} {\bibfnamefont {B.}~\bibnamefont
  {Shapiro}},\ }\bibfield  {title} {\bibinfo {title} {{Arbitrary steering of
  multiple particles independently in an electro-osmotically driven
  microfluidic system}},\ }\href {https://doi.org/10.1109/TCST.2006.876636}
  {\bibfield  {journal} {\bibinfo  {journal} {IEEE Transactions on Control
  Systems Technology}\ }\textbf {\bibinfo {volume} {14}},\ \bibinfo {pages}
  {669} (\bibinfo {year} {2006})}\BibitemShut {NoStop}%
\bibitem [{\citenamefont {Khalil}(2002)}]{Khalil2002}%
  \BibitemOpen
  \bibfield  {author} {\bibinfo {author} {\bibfnamefont {H.~K.}\ \bibnamefont
  {Khalil}},\ }\href@noop {} {\emph {\bibinfo {title} {Nonlinear systems; 3rd
  ed.}}}\ (\bibinfo  {publisher} {Prentice-Hall},\ \bibinfo {address} {Upper
  Saddle River, NJ},\ \bibinfo {year} {2002})\BibitemShut {NoStop}%
\bibitem [{\citenamefont {Yang}\ \emph {et~al.}(2014)\citenamefont {Yang},
  \citenamefont {Jiang},\ and\ \citenamefont {Cocquempot}}]{Yang2014}%
  \BibitemOpen
  \bibfield  {author} {\bibinfo {author} {\bibfnamefont {H.}~\bibnamefont
  {Yang}}, \bibinfo {author} {\bibfnamefont {B.}~\bibnamefont {Jiang}},\ and\
  \bibinfo {author} {\bibfnamefont {V.}~\bibnamefont {Cocquempot}},\ }\bibfield
   {title} {\bibinfo {title} {{A survey of results and perspectives on
  stabilization of switched nonlinear systems with unstable modes}},\ }\href
  {https://doi.org/10.1016/j.nahs.2013.12.005} {\bibfield  {journal} {\bibinfo
  {journal} {Nonlinear Analysis: Hybrid Systems}\ }\textbf {\bibinfo {volume}
  {13}},\ \bibinfo {pages} {45} (\bibinfo {year} {2014})}\BibitemShut {NoStop}%
\bibitem [{\citenamefont {M{\"{u}}ller}\ and\ \citenamefont
  {Liberzon}(2012)}]{Muller2012}%
  \BibitemOpen
  \bibfield  {author} {\bibinfo {author} {\bibfnamefont {M.~A.}\ \bibnamefont
  {M{\"{u}}ller}}\ and\ \bibinfo {author} {\bibfnamefont {D.}~\bibnamefont
  {Liberzon}},\ }\bibfield  {title} {\bibinfo {title} {{Input/output-to-state
  stability and state-norm estimators for switched nonlinear systems}},\ }\href
  {https://doi.org/10.1016/j.automatica.2012.06.026} {\bibfield  {journal}
  {\bibinfo  {journal} {Automatica}\ }\textbf {\bibinfo {volume} {48}},\
  \bibinfo {pages} {2029} (\bibinfo {year} {2012})}\BibitemShut {NoStop}%
\bibitem [{\citenamefont {Huang}\ \emph {et~al.}(2017)\citenamefont {Huang},
  \citenamefont {Schofield},\ and\ \citenamefont {Kapral}}]{Huang2017}%
  \BibitemOpen
  \bibfield  {author} {\bibinfo {author} {\bibfnamefont {M.-J.}\ \bibnamefont
  {Huang}}, \bibinfo {author} {\bibfnamefont {J.}~\bibnamefont {Schofield}},\
  and\ \bibinfo {author} {\bibfnamefont {R.}~\bibnamefont {Kapral}},\
  }\bibfield  {title} {\bibinfo {title} {{Chemotactic and hydrodynamic effects
  on collective dynamics of self-diffusiophoretic Janus motors}},\ }\href
  {https://doi.org/10.1088/1367-2630/aa958c} {\bibfield  {journal} {\bibinfo
  {journal} {New Journal of Physics}\ }\textbf {\bibinfo {volume} {19}},\
  \bibinfo {pages} {125003} (\bibinfo {year} {2017})}\BibitemShut {NoStop}%
\end{thebibliography}%

\end{document}


\title{Supplemental Information for ``Chemical herding as a multiplicative factor for top-down manipulation of colloids ''}

\maketitle

\section{Gradient vector field }

In this section, we give the equations for the gradient vector field (GVF) used to set the trajectory of the herder. A GVF is a function $\bm{V}_{g}(\bm{y},\bm{y}^{*},\{\bm{x}_{i}\})$ that produces a direction for the herder to move at each time step. (Recall that $\bm{y}$ is the position of the herder, $\bm{y}^*$ is the target position for the herder, and $\{\bm{x}_{i}\}$ is the set of the positions of each follower.) We use a modified version of the GVF presented by Wilhelm and Clem 
to calculate the direction $\bm{V}_g/\tilde{V}_g$ that will move the herder to $\bm{y}^{*}$ while avoiding collisions with the follower particles.

The GVF is described by  
\begin{equation}
\label{eq-GVF}
    \bm{V}_{g} = \bm{V}_{\mathrm{path}} + \sum_{i=0}^{n_{\mathrm{f}}} P_{i} \bm{V}_{\mathrm{obs},i},
\end{equation}
where $\bm{V}_{\mathrm{path}} $ is the 2D vector that guides the herder to its optimal placement $y^{*}$, and $\bm{V}_{\mathrm{obs},i}$ is the vector that repels the herder from obstacle $i$. 
Also, $P_{i}$ is a decay function that determines how far from obstacle $i$ the repulsion persists, defined as 
\begin{equation}
\label{eq-GVFdecay}
    P_{i} = -\tanh \left(\frac{2\pi \tilde{d}_{ih}}{R_i} -\pi \right)+1,
\end{equation}
where  $\tilde{d}_{ih}= ||\bm{y}-\bm{x}_{i}||$ (the distance between obstacle $i$ and the herder). For followers, we set $R_i = kR_{fh}$, where the value of $k = 1.6$ was chosen by trial and error to ensure the particles do not overlap.
In simulations with multiple herders, we set a different value of $R_i$ for herders to prevent them from moving too close to each other, settling on a value of ${R_i= 16 \mu}$m.

Each of the terms in Equation~\eqref{eq-GVF} is further divided into a convergence term and a circulation term. 
The convergence terms make the herder either approach its target or avoid an obstacle. 
The circulation terms help the GVF avoid singularities, or points where $||\bm{V}_g|| = 0$. 
Each term is weighted appropriately, giving
\begin{align}
    \bm{V}_{\mathrm{path}} = G_{\mathrm{path}} \bm{V}_{\mathrm{path}}^{\mathrm{conv}} + H_{\mathrm{path}} \bm{V}_{\mathrm{path}}^{\mathrm{circ}}\\
    \bm{V}_{\mathrm{obs},i} = G_{\mathrm{obs}} \bm{V}_{\mathrm{obs},i}^{\mathrm{conv}} + H_{\mathrm{obs}} \bm{V}_{\mathrm{obs,},i}^{\mathrm{circ}}
\end{align}
which introduces the functions 
$\bm{V}_{\mathrm{path}}^{\mathrm{conv}}$,$\bm{V}_{\mathrm{path}}^{\mathrm{circ}}$,$\bm{V}_{\mathrm{obs}}^{\mathrm{conv}}$, and $\bm{V}_{\mathrm{obs}}^{\mathrm{circ}}$ that will be defined in the following paragraphs, as well as a set of constant weights.  
To calculate $\bm{V}_{\mathrm{path}}$, we used a convergence weight $G_{\mathrm{path}}=1$ and a circulation weight ${H_{\mathrm{path}}=0}$, meaning there is no circulation when the herder is far from any obstacle. 
To calculate each $\bm{V}_{\mathrm{obs},i}$ we used a convergence weight $G_{\mathrm{obs}} = -2$, meaning the herder will be repelled from obstacles, and a circulation weight ${H_{\mathrm{obs}} = 0.5}$, which will bias the particle into clockwise movement around obstacles. 
These parameters were selected by trial and error.

The convergence and circulation terms of the portion of the GVF that make the herder approach its goal are
\begin{equation}
    \bm{V}_{\mathrm{path}}^{\mathrm{conv}} = \frac{-1}{\sqrt{(\cos(\delta)x + \sin(\delta)y)^2}} 
    \begin{bmatrix}
        x\cos^2(\delta) + \cos(\delta)\sin(\delta)y \\ y\sin^2(\delta) + \cos(\delta)\sin(\delta)x
    \end{bmatrix}
\end{equation}
and
\begin{equation}
    \bm{V}_{\mathrm{path}}^{\mathrm{circ}} = [\sin(\delta), -\cos(\delta)]^T,
\end{equation}
where $\delta$ is the angle between the desired path and the $x$-axis.

The convergence and circulation terms that prevent the herder from colliding with an obstacle centered at $(x_c,y_c)$ are 
\begin{equation}
    \bm{V}_{\mathrm{obs}}^{\mathrm{conv}} = \frac{-1}{\bar{x}^2 + \bar{y}^2} 
    \begin{bmatrix}
        2\bar{x}^3 + 2\bar{x}\bar{y}^2 \\
        2\bar{y}^3+2\bar{x}^2\bar{y}
    \end{bmatrix}
\end{equation}
and
\begin{equation}
    \bm{V}_{\mathrm{obs}}^{\mathrm{circ}} = [2\bar{y} -2\bar{x}]^T,
\end{equation}
where $\bar{x} = x - x_c$ and $\bar{y} = y - y_c$.

The GVF, as given so far, encounters a problem when the followers are close together. 
When an unchased follower is too close to the chased follower, treating that unchased follower as an obstacle may prevent the herder from ever reaching its optimal placement $\bm{y}^*$. 
Our heuristic for solving this problem is that we consider only those followers that are located further than $R_{fh}$ from $\bm{y}^{*}$ as obstacles to be included in the sum in Equation~\eqref{eq-GVF}. 
Collisions with obstacles closer than that are deemed necessary.

In Equation~\eqref{eq-GVF}, $\bm{V}_g$ gives the direction for the herder to move. 
We would also like the herder to move at the maximum speed $v_{\mathrm{max}}$ allowed by the physical system. 
However, since the controller is discrete in time, care must be taken to ensure the herder does not overshoot its target. 
Thus the force $\bm{F}_{\mathrm{ext,h}}$ applied to the herder dynamics is set to be
\begin{equation}
\label{eq-GVFveloc}
    \bm{F}_{\mathrm{ext,h}} = \gamma_{h}\frac{\bm{V}_{g}}{\tilde{V}_{g}} \min \left(v_{\mathrm{max}},\frac{\tilde{e}_{h}}{\Delta t_{\mathrm{control}} }\right),
\end{equation}
where $\tilde{e}_{h} = ||\bm{y} - \bm{y}^{*}||$ is the vector from the herder to its target position and $\Delta t_{\mathrm{control}}$ is the timestep of the controller. 
This velocity ensures that the herder moves at a speed of $v_{\mathrm{max}}$ until it is within the distance from its optimal position that it can move in a single timestep, then it slows to avoid overshooting. 
We used a value of $\Delta t_{\mathrm{control}}=0.1$ s.

\section{Pseudo-steady state approximation}

In this section, we use dimensional analysis to show that a pseudo-steady state approximation can be used to find the concentration field around a herder. 
We start with the reaction-diffusion equation
\begin{equation}
\label{eq-reactiondiffusion}
    \frac{\partial C_s}{\partial t} = D_s \nabla^2 C_s + g_h \delta(\bm{X}-\bm{y}).
\end{equation}
We then nondimensionalize using $\tilde{\bm{X}} = \bm{X}/R_{fh}$, $\tilde{\bm{y}} = \bm{y}/R_{fh}$, $\tilde{C_s} = C_s/C_{\infty}$, and $\tilde{t} = t/t^*$. For the time scale $t^*$, we use the length of time it takes the herder to move a chased particle a distance of $R_{fh}$, which, from Equation~39 of the main paper, is  given by
$t^*=R_{fh}^3/k_{\mathrm{diff}}$.  Then, substituting in the dimensionless variables, we have
\begin{equation}
    \frac{R_{fh}^2}{D_s t^*} \frac{\partial \tilde{C_s}}{\partial \tilde{t}} = \tilde{\nabla}^2 \tilde{C_s} + \frac{g_h R_{fh}}{D_s C_{\infty}} \delta(\tilde{\bm{X}} - \tilde{\bm{y}}).
\end{equation}
If the Peclet number $Pe = R_{fh}^2/D_st^*$ is small, then diffusion will dominate over the motion of the herder, and the $\partial C_s/\partial t$ term of Equation~\eqref{eq-reactiondiffusion} will be negligible. 
Using the values of parameters given in the main paper, we have $Pe \approx 2\times 10^{-3}$. Since this is much smaller than unity, we conclude that a pseudo-steady state approximation is valid. We have also verified the validity of the pseudo-steady state solution numerically to our satisfaction.